\documentclass[
a4paper,
showpacs,
twocolumn,
superscriptaddress,reprint,amsmath,amssymb,aps,prb,longbibliography,showpacks]{revtex4-2}
\usepackage{amssymb}
\usepackage{amsmath}
\usepackage{amsfonts}
\usepackage{braket}
\usepackage{graphicx}
\usepackage{bm}
\usepackage{color}
\usepackage{multirow}
\usepackage{natbib}
\usepackage{hyperref}
\usepackage[normalem]{ulem}
\usepackage{verbatim}
\usepackage{dcolumn}
\usepackage{ulem}
\usepackage{float}

\usepackage[dvipsnames]{xcolor}

\def \be {\begin{equation}}
\def \ee {\end{equation}}
\def \bea {\begin{align}}
\def \eea {\end{align}}
\def \p {\partial}
\def \BEA {\begin{eqnarray}}
\def \EEA {\end{eqnarray}}

\def \BC {\begin{cases}}
\def \EC {\end{cases}}

\newcommand{\aleq}[1]{
\begin{equation}
 \begin{aligned}
 #1
 \end{aligned}
\end{equation}
}

\usepackage[T2A]{fontenc}
\usepackage{hyphenat}%
\usepackage[english]{babel}

\begin{document}
\title
{ 
Ratchet effect in lateral plasmonic crystal: Giant enhancement due to interference of ``bright'' and ``dark'' modes} 

\author{I.\,V.~Gorbenko}
\address{Ioffe Institute,
194021 St.~Petersburg, Russia}
\author{S.\,O.~Potashin}
\address{Ioffe Institute,
194021 St.~Petersburg, Russia}
\author{V.\,Yu.~Kachorovskii }
\address{Ioffe Institute,
194021 St.~Petersburg, Russia}

%\date{\today}
%\pacs{72.80.Vp, 73.23.Ad, 73.63.Bd}

\begin{abstract} 
We develop a theory of the ratchet effect in a lateral plasmonic crystal (LPC) formed by a two-dimensional electron gas under a periodic dual-grating gate. The system is driven by terahertz radiation, and the spatial asymmetry required for the generation of dc photocurrent is introduced by a phase shift between the radiation’s near-field modulation and the static electron density profile. In contrast to the commonly used perturbative ``minimal model'' of the ratchet effect, which assumes weak density modulation, we solve the problem exactly with respect to the static gate‑induced potential while treating the radiation field perturbatively. This approach reveals a dramatic enhancement of the plasmonic contribution to the ratchet current due to the interference of “bright” and “dark” plasmon modes, which are excited on an equal footing in the asymmetric LPC. Specifically, we predict a parametric growth of the plasmonic peak as compared with the Drude peak with increasing coupling, and the appearance of a dense super‑resonant structure when the spacing between plasmonic sub‑bands becomes larger than the damping rate. Hence, the dc response exhibits both resonant and super‑resonant regimes observed in recent experiments on the radiation transmission through the LPC. The interplay of bright and dark modes, together with their interference, provides a powerful mechanism for controlling the magnitude and sign of the photocurrent by gate voltages and the radiation frequency.

\end{abstract}

\maketitle

\section{Introduction}

The generation of direct electric current (dc) by external electromagnetic radiation represents a fundamental physical process in optoelectronics. The emergence of a non-zero dc photoresponse requires the simultaneous fulfillment of two essential conditions:
 (i) nonlinearity within the system, which is necessary to rectify the incoming alternating current (ac) signal; (ii) broken inversion symmetry, which provides the structural asymmetry needed to define a preferential direction for the generated dc current.
Provided both of the above conditions are satisfied, a dc current arises due to the so-called ratchet effect, 
which has a rather universal nature and takes place in various systems (for a review, see \cite{Linke2002, Reimann2002, Haenggi2009, Ivchenko2011, Denisov2014, Cubero2016, Budkin2016a, Reichhardt2017, Ganichev2017}).

In crystalline systems lacking inversion symmetry at the lattice level, the ratchet effect is inherently microscopic (see, e.g., \cite{Drexler2013}). In recent decades, however, artificially engineered periodic structures without inversion symmetry have attracted growing attention. Their key advantage lies in enabling full electrical control of the ratchet effect.
For instance, a direct-current (dc) ratchet photocurrent can be generated in an optically excited two-dimensional electron gas when its electron density is spatially modulated in an asymmetric manner. This is achieved using a metal grating gate with two sub-gratings—referred to as a dual grating gate structure (DGGS). By adjusting the relative voltage between the two sub-gratings of the DGGS, one can precisely control both the magnitude and the sign of the resulting photocurrent \cite{Faltermeier2015,Faltermeier2017,Faltermeier2018,Hubmann2020,Sai2021,Monch2023}.

Interest in such artificial systems is motivated by both fundamental and applied considerations. These structures are of particular significance for two-dimensional (2D) plasmonics, a field that originated several decades ago (for early work, see \cite{chap1972,allen1977,Theis1977a, Tsui1978, Theis1978, Theis1980, Tsui1980a}; for a review, see \cite{Maier2007}).

Since the pioneering experiments of Allen, Tsui, and co-authors \cite{allen1977,Theis1977a, Tsui1978, Theis1978, Theis1980, Tsui1980a}, a standard method for coupling light to plasma waves in a 2D system has been the use of a metallic grating coupler. Placed between the light source and the conducting channel, this grating acts as an antenna, modulating the two-dimensional electron gas with a wave vector $k=2\pi/L$, where $L$ is the grating period, thereby exciting plasmons of the same wave vector. By applying a gate voltage $U_{\rm g}$ to the coupler, one can vary the plasma wave velocity $s =\sqrt{e (U_{\rm g} -U_{\rm th})/m}$ (where $U_{\rm th}$ is the threshold voltage and, consequently, control the plasmon frequency $\omega=s k$. This tuning mechanism operates in single-gated structures, such as field-effect transistors, as well as in more complex multi-gated devices.

For a typical period $L$ on the order of several microns and a typical plasmon velocity $s \sim 10^8$ cm/s, the plasmon oscillation frequency  falls into the terahertz (THz) range. This frequency can be tuned over a broad interval by adjusting the gate voltage.

Relevant applications are found in terahertz nanoelectronics,which is focused on developing compact, resonant, and gate-tunable plasmonic THz emitters and detectors. These include devices based on single field-effect transistors (see \cite{Dyakonov1993,Dyakonov1996, Vicarelli2012, Kachorovskii2013, Ikamas2018} and the reviews such as  \cite{Stillman2007}) and those utilizing grating-gate structures (see \cite{Boubanga-Tombet2014} and reviews \cite{Ganichev2017, Lau2020, Otsuji2021}).

High-quality plasmonic resonances require high-mobility 2D systems, with recent emphasis on novel graphene-based materials \cite{Grigorenko2012, Vicarelli2012,DiPietro2013,Kachorovskii2013,Elkhatib2011,Giorgianni2016,Rumyantsev2015,Autore2017,Politano2017,Yang2018,Bandurin2018, Otsuji2021,Otsuji2022}. Gate-tunable plasmonic resonances with high quality factors were first demonstrated in the transmission coefficient of two-dimensional multi-gate lateral superlattices based on GaN/AlGaN heterostructures \cite{Muravjov2010TemperatureStructures} at temperatures up to 170 K, and more  recently at room temperature \cite{Boubanga-Tombet2020} in graphene-based structures.

The ratchet effect in multi-gated structures has also been studied extensively, both in zero magnetic field (see \cite{Ivchenko2011,Popov2011,Rozhansky2015,Faltermeier2017,Faltermeier2018,Hubmann2020,Sai2021,Monch2022,Monch2023} and references therein) and in sufficiently weak magnetic fields \cite{Faltermeier2017,Faltermeier2018,Hubmann2020,Sai2021,Monch2022,Monch2023}. A plasmonic splitting of the cyclotron resonance in the ratchet photocurrent has been observed and described theoretically \cite{Monch2023}. Furthermore, the ratchet effect in grating-gate structures has proven to be a powerful tool for the optical probing the hydrodynamic regime of the electron liquid \cite{Monch2022,Monch2023, Potashin2025}.

Despite considerable research on multi-gate systems, several important aspects remain unexplored. A particularly promising direction is the use of 2D multi-gate configurations as tunable lateral plasmonic crystals (LPCs). Their unique appeal lies in the ability to engineer gate-tunable allowed and forbidden bands for plasma waves.

The concept of a tunable LPC was proposed over a decade ago \cite{Kachorovskii2012} and later investigated theoretically \cite{Petrov2017,Aizin2023}. It received experimental confirmation in recent work \cite{Sai2023}, which reported tunable plasmonic resonances in the transmission of a GaN/AlGaN-based LPC and demonstrated the transition  between weak and strong coupling regimes, in good agreement with the theory further developed in Ref.~\cite{Gorbenko2024}. Subsequent studies \cite{Dub2024, Khisameeva2025} experimentally investigated the dependence of plasmonic resonances on gate voltage, wave vector, and filling factor, again finding good agreement with theoretical predictions.

The purpose of the present work is to develop a theory of the ratchet effect in a lateral plasmonic crystal.

\color{black}
 \section{Problem formulation} \label{Sec-approach}
 \subsection{Minimal purturbative model}
Existing   works on the analytical theory of the ratchet effect in grating-gate structures \cite{Olbrich2009, Olbrich2011, Ivchenko2011, Rozhansky2015, Olbrich2016, Faltermeier2018, Sai2021, Monch2022, Monch2023, Potashin2025} are based on the very simplified minimal perturbative model proposed in \cite{Olbrich2009, Ivchenko2011}.
Within this model, it is assumed that a grating gate, to which a voltage is applied, creates a weak periodic static potential $U_g(x)$ in the 2D channel and also modulates the amplitude of the electromagnetic near-field acting in the 2D channel, which is also assumed to be weak. Furthermore, only the zeroth and first harmonics are taken into account for both the potential and the field:
\be
U_g(x)= U_g+U_0 \cos k x
\label{Eq-Ug_weak}
\ee
\begin{equation}
 E(x,t) = E_0[1+ h \cos(k x + \phi)]\cos \omega t , 
 \label{Eq-grating_modulation}
\end{equation}
where $E_0$ is the amplitude of the incoming wave and $h$ is the modulation depth: $U_0 \ll U_g, h \ll 1.$ 

The key ingredient of the minimal model is the phase shift $\phi.$ For $\phi=\pi n,$ where $n$ is integer, there is inversion center in the system of Eqs.~\eqref{Eq-Ug_weak} and \eqref{Eq-grating_modulation} and generation of dc current is prohibited by the symmetry. For $\phi \neq \pi n,$ inversion center is absent, so that the grating gate structure is asymmetric thus allowing for generation of non-zero dc response, $J_{\rm dc } \propto \Xi,$ which is proportional to the asymmetry parameter \cite{Ivchenko2011}
\begin{equation} 
\Xi = \left< E(x,t)^2 \frac{\partial U(x)}{\partial x} \right>_{x,t}= \frac{h k E_0^2 U_0}{{4}} \sin{\phi},
\label{Eq-asymmetry}
\end{equation}
It is clear from this formula that a nonzero dc response is obtained in the third order of the perturbative analysis of the problem (with respect to $U_0$ and $E_0$): $J \propto E_0^2 U_0.$

The minimal model
has proven to be highly effective and has allowed for the description of key features in a number of experiments on radiation-to-dc current conversion \cite{Olbrich2011, Ivchenko2011, Olbrich2016, Faltermeier2017,Faltermeier2018,Hubmann2020, Sai2021,Monch2022,Monch2023}, as well as for the study of the transition to the hydrodynamic regime of a 2D electron fluid \cite{Rozhansky2015,Monch2022}. At the same time, the perturbative analysis that underlies this model is generally not applicable to realistic LPC where density modulation can be sufficiently strong.

The aim of the present work is to develop a  non-perturbative theory of the ratchet effect applicable to the case of strong density modulation and to investigate the transition between weak and strong coupling regimes of LPC controlled by density modulation  depth.  The key advancement compared to the minimal model is that the static potential modulating the electron density in the 2D channel is no longer considered small; that is, the problem is solved exactly with respect to $U_0$, but still perturbatively with respect to $E_0$.

Below we develop  such a theory and obtain  some  new results, which are not captured by the minimal model:
\begin{itemize}
    \item 
 with increasing the density modulation, the so-called ``dark'' modes with sufficiently high amplitudes (approximately the same as for ``bright'' modes) become visible in the excitation spectra; 
\item  the plasmonic resonances become strongly asymmetric with the shape of the Fano type; \item  the amplitudes of the Drude and plasmonic peaks become parametrically different. Specifically, interference of two types of modes leads to giant (one- two orders of magnitude) enhancement of the fundamental plasmonic resonance. 
\end{itemize}
We also demonstrate that at not too small coupling the dc photoresponse shows both resonant and super-resonant regimes observed \cite{Sai2023} and explained \cite{Gorbenko2024} for the transmission coefficient through LPC. For fixed coupling (i.e. fixed modulation depth) transition between two regimes is controlled by the quality factor of plasmonic resonances.

\subsection{Non-perturbative approach}
In this paper, we consider DGGS, i.e. a two-dimensional electron gas in the $(x,y)$ plane, placed in a grating-gate potential.  Instead of harmonic periodic modulation, we consider a potential   which is uniform in the $y$ direction and periodic in the $x$ direction with the period $L=L_1+L_2$, having  a step-like  form within the unit cell: 
\be
U_{\rm g}(x) =\left\{ \begin{array}{ll}
 \, U_{\rm g}^{(1)}, ~  \, \, \, {\rm for} \, \, \, 0 <x<L_1 \quad\text{(region~ 1)},
 \\
 \, U_{\rm g}^{(2)}, \quad \, \, \, {\rm for} \, \, \,L_1<x<  L_1+L_2 ~~\text{(region~ 2)}.
\end{array} \right.
\label{Eq-Ug}
\ee
This model describes the realistic experimental  
DGGS-based setups much better than the minimal model.

The potential $U_g(x)$ creates a LPC with allowed and forbidden bands related to propagation in $x$ direction \cite{Kachorovskii2012,Gorbenko2024}.
 The inverse lattice vector of LPC is given by
\begin{equation}
k=\frac{2\pi}{L_1+L_2}.
\label{k-vector}
\end{equation}
Gate voltage $U_{\rm g}(x)$ controls stationary electron concentration $N$ in the channel and plasma wave velocity: $N=N_1, ~s=s_1$ in the region ``1'' and $N=N_2,~ s=s_2,$ in the region ``2'', 
where
 \begin{equation}
 \begin{aligned}
 &N_{1,2}=\frac{C \left [U_g^{(1,2)}- U_{\rm th}\right ]}{e}=\frac{C \, U_{1,2}}{e}, 
 \\
&s_{1,2}=\sqrt{\frac{e^2 N_{1,2} }{m C}}.
 \end{aligned}
 \label{Eq-N0}
\end{equation}
 Here $C=\epsilon/4\pi d $ is the channel capacitance per unit area, $d$ is the spacer width, and $\epsilon$ is the dielectric constant (for simplicity, we assume $\epsilon$ is constant throughout the system), $U_{\rm th}$ is the so-called threshold voltage, and $U_{1,2}=U_g^{(1,2)}- U_{\rm th}$. 
To be specific, we assume that $s_1 \geq s_2$.  Below, we fix $U_g^{(1)}$  (and, consequently, $N_1$ and $s_1$)  assuming  that $N_2$ and $s_2$ are tuned by $U_g^{(2)},$ which changes in the interval $0<U_g^{(2)}< U_g^{(1)}$ ($0<s_2<s_1$).
Weak coupling regime corresponds to almost homogeneous grating-gate potential, so that $s_1-s_2 \ll s_1.$ Strong coupling regime is realized when passive regions are depleted by sub-grating ``2'', so that $s_2 \ll s_1$. 
Following Refs.\cite{Boubanga-Tombet2020, Gorbenko2024}, we designate region ``1'' as {\it active} and region ``2'' as {\it passive}. Physics behind this designation is related to the strong coupling regime, when plasma waves in regions ``2'' have a relatively high damping rate, so that different regions ``1'' are isolated with respect to ac current, which decays exponentially into depleted regions ``2'' within a narrow boundary layers (see a more detailed discussion in Ref.~\cite{Boubanga-Tombet2020}).

 We describe coupling by the parameter 
\be
\Delta = \frac{s_1-s_2}{s_1},
\ee
 which is controlled by $U_g^{(2)}$ and changes from $\Delta \ll 1 $ in the weak coupling regime to $\Delta \approx 1$ in the strong coupling regime.

 We consider dc photoresponse of the DGGS to the electromagnetic linearly-polarized (along $x$ axis) radiation with frequency $\omega,$ with the corresponding wavelength, $\lambda$ much larger than period of grating $L = L_1+L_2$ (in particular, for terahertz radiation $\lambda$ is about hundred microns, while typical $L$ is about several microns). 
 The most general form of the radiation-induced near-field in the 2D channel reads 
\be
E(x,t) = E_0 [1+ \sum_{n=1}^{n=\infty} h_n \cos(n k x + \phi_n)] \cos \omega t,
\label{Eq-grating}
\ee
where $E_0$ is homogeneous component, $h_n $ and $\phi_n$ are the modulation strength and the phase of $n-$th near-field harmonic, respectively. 

The coefficients $h_n$ are determined with geometry of grating and usually found numerically. Following Ref.~\cite{Ivchenko2011}, we assume $ h_{n\neq 1} \ll h_1$ and keep the first harmonic only, where we put $h_1=h, \phi_1=\phi.$ thus arriving at Eq.~\eqref{Eq-grating_modulation}.
 It is worth noting that inequality $ h_{n\neq 1} \ll h_1,$ while not proven analytically,
 is supported by numerical analysis \cite{Petrov2014} (see, also, discussion of near-field effects in Refs. \cite{Gorbenko2024, Gorbenko2025}).
 
Thus, we use the same expression for the electric field as in the minimal model, but unlike that model, we assume that the modulating potential and consequently the density modulation depth can be arbitrarily large. The asymmetry of the system deserves a special comment. Although static potential \eqref{Eq-Ug} has inversion center, the total potential including also the potential of the external electrical does not have inversion center provided that $\phi \neq \phi_{\rm sym} = - \pi L_1/L.$ 
The asymmetry parameter for the PC  can be determined as follows:
\begin{equation} 
\begin{aligned}
\Xi_{\rm PC} & = h E_0^2 \left(\frac{U_{\rm 1}^2-U_{\rm 2}^2}{U_{\rm 1}^2+U_{\rm 2}^2}\right) \sin(\phi -\phi_{\rm sym} ).
\end{aligned}
\label{Eq-asym_PC}
\end{equation}
Comparing with Eq.~\eqref{Eq-asymmetry}, we see that  there is  a shift $\phi_{\rm sym}$  which arises because  the position of the inversion center of the density modulation \eqref{Eq-Ug} is given by $x=L_1/2$ in contrast to potential Eq.~\eqref{Eq-Ug_weak} which is symmetric with respect to $x=0.$ For $L_1=L_2=L/2,$ we get $\phi_{\rm sym} = -\pi/2,$
so that $\sin(\phi-\phi_{\rm sym})=  \cos \phi. $ 
% The relation with previous work where the parameter $\sin{\theta}$ was used (for example, in \cite{Rozhansky2015, Monch2022}) is evident: these angles are connected as, $\phi = \theta - \pi/2$.

As we show below, in the weak coupling regime, when  $\Delta \ll 1$ the results of the minimal model are reproduced within our model. At the same time, in the cases of strong ($1-\Delta \ll 1$) and intermediate ($\Delta \sim 1$) coupling the results differ significantly.

\subsection{Hydrodynamic approximation}
We use hydrodynamic approach assuming that the electron-electron collision time is shorter than the electron-impurity and electron-phonon scattering times. Then, dynamics of $n_{\alpha}$ and the electron drift velocity, $v_{\alpha},$ obey standard hydrodynamic equations ($\alpha=1,2$):
\begin{align}
&\frac{\partial {v_{\alpha}}}{\partial t}\! + \!{v_{\alpha}} \frac{\p {v_{\alpha}}}{\p x} \!+\! \gamma {v_{\alpha}}\!=\! - s_{\alpha}^2 \frac{\p n_{\alpha}}{\p x} + \frac{{F}(x, t)}{m},
\label{Eq_Navier_Stokes}
\\ 
&\frac{\partial n_{\alpha}}{\partial t} + \frac{ \p [(1+ n_{\alpha}) {v_{\alpha}}]}{\p x} = 0. 
\label{Eq_continuity}
\end{align}
Here,  $\gamma$ is inverse momentum relaxation and we neglect relaxation related to the viscosity of the electron liquid, $\eta,$ by assuming that $\eta q^2 \ll \gamma$ for typical wave vectors $q.$ 
We took into account in Eqs.~\eqref{Eq_Navier_Stokes} and \eqref{Eq_continuity}
 that drift velocity is parallel to $x$ axis for linearly polarized radiation. 

A  comment is needed about r.h.s of Eq.~\eqref{Eq_Navier_Stokes}.
The total oscillating force, $F_{\rm tot} (x,t)$, acting on the electron in the 2D channel is directed along axis $x$ and is given by the sum of the the external force from the linear polarized wave, $F= e E(x, t)$ , and plasmonic field: $$F_{\rm tot}(x,t) = F - {e} \frac{\p U}{\p x},$$ 
where $F=e E(x, t)$ and 
$U=U(x,t)$ is the local voltage swing between 2D channel and gate electrode: $$U(x,t) = U_g(x)+\delta U(x,t).$$  Here, $\delta U(x,t)$ is plasmonic potential created by radiation-induced inhomogeneous density oscillations. This potential is different in regions ``1'' and ``2''. Introducing dimensionless concentrations:
$$n_{i}=\frac{N(x,t) -N_i}{N_i},\quad i=(1,2) $$
(here $N_i$ are given by Eq.~\eqref{Eq-N0}), we get 
\be
\frac{e}{m}\frac{\p \delta U_{\alpha}}{\p x} = s_{\alpha}^2 \frac{\p n_{\alpha}}{\p x}.
\label{Eq_delta_U}
\ee
To solve equations \eqref{Eq_Navier_Stokes} and \eqref{Eq_continuity} we use boundary conditions between regions ``1'' and ``2'', that correspond to the continuity of total current $J^{\rm tot}=e N v \propto s^2 (1+n) v ,$ with $J^{\rm tot} = J + j$ where $J$ is dc current and $j$ is ac current, and variable $\epsilon = s^2 n+v^2/2, $ which is proportional to the electron energy in the channel \cite{Kachorovskii2012, Petrov2017}:
\begin{equation}
J_1^{\rm tot}=J_2^{\rm tot},\quad \epsilon _{1}=\epsilon_{2}, 
 \label{Eq-BC}
\end{equation}
at the boundary between regions.

\section{Solution}
\label{Sec-solution}
We will search solution of 
Eqs.\eqref{Eq_Navier_Stokes} and \eqref{Eq_continuity} perturbatively with respect to $E_0$:
\aleq{
&n_\alpha(x,t) = 
 n_\alpha(x) e^{-i \omega t} + n_\alpha^*(x) e^{i \omega t}+ \delta n_\alpha(x),
\label{Eq_n}
\\
&v_\alpha(x,t) = 
 v_\alpha(x) e^{-i \omega t} + v_\alpha^*(x) e^{i \omega t}+ \delta v_\alpha(x),
%\label{Eq_v}
}
where ${n_\alpha, v_\alpha} \propto E_0$ describe linear response, ${\delta n_\alpha, \delta v_\alpha} \propto E_0^2$ are rectified time-independent  concentration and velocity that appear due to non-linearity of the system, and $\alpha = (1, 2)$. 

Next two steps are as follows: we first linearize Eqs.~\eqref{Eq_Navier_Stokes}, \eqref{Eq_continuity} and find $n_\alpha, v_\alpha,$ and then substitute these solutions to the non-linear terms of Eqs.\eqref{Eq_Navier_Stokes}, \eqref{Eq_continuity} and find $\delta n_\alpha$ and $\delta v_\alpha.$ 

\subsection{ Linearization of hydrodynamic equations}
Linearization of Eqs.~\eqref{Eq_Navier_Stokes}, \eqref{Eq_continuity} yields a system of coupled equations for the amplitudes $n_\alpha(x)$ and $v_\alpha(x)$: 
\begin{align}
&-i \omega v_\alpha + \! \gamma v_\alpha + s_\alpha^2 \frac{\partial n_\alpha}{\partial x}= \frac{eE_0[1+ h \cos(k x + \phi)]}{m},
\label{Eq-Navier_Stokes_1}
\\ 
&- i \omega n_\alpha + \frac{\partial v_\alpha}{\partial x} = 0. 
\label{Eq-continuity_1}
\end{align}
 Solution of these equations can be written as
\be
\begin{aligned}
&n_\alpha =A_\alpha e^{i q_\alpha x}+B_\alpha e^{-i q_\alpha x}+n^{\rm ext}_\alpha, 
\\
&v_\alpha=(A_\alpha e^{i q_\alpha x}-B_\alpha e^{-i q_\alpha x})\frac{\omega}{q_\alpha}+ v^{\rm ext}_\alpha.
\end{aligned}
\label{vxy}
\ee
Here
\be
q_\alpha = \frac{\sqrt{\omega (\omega+i \gamma) }}{s_\alpha}
%\frac{\Omega+ i \Gamma}{s_\alpha}
\label{Eq-wv}
\ee
are the complex wave vectors,
\be
\begin{aligned}
&n_\alpha^{\rm ext}=-h \frac{e E_0 k}{2 m s_\alpha^2(q_\alpha^2-k^2)} \sin{(k x+\phi)},
\\
&v_\alpha^{\rm ext}=\frac{i e E_0}{2 m (\omega+i \gamma)} + h \frac{i \omega e E_0 }{2 m s_\alpha^2(q_\alpha^2 - k^2)} \cos{(k x+\phi)},
\end{aligned}
\label{Eq_nv_ext}
\ee
are the concentration and velocity that would be induced by external radiation in the infinite system with plasma velocity $s_\alpha.$ The time- and coordinate-independent coefficients 
$A_\alpha$ and $B_\alpha$ arise due to coupling between regions ``1'' and ``2'' and should be found from the boundary conditions, which can be obtained by linearization of   Eq.~\eqref{Eq-BC}:
\be
j \propto s_1^2 v_1=s_2^2 v_2,\quad s_1^2 n_1=s_2^2 n_2.
\label{Eq-BC1}
\ee

\subsection{Bright and dark modes} \label{bright-dark}
Before performing calculations of the linear response to an external electromagnetic field, let us recall a number of properties of plasma excitations in LPC in the absence of illumination, for $E_0=0$.
For zero plasma wave damping, $\gamma=0,$
 spectrum of the LPC, $\omega (K), $ obeys \cite{Kachorovskii2012}:

\begin{align} \label{Eq-natural_frequency}
&\cos{\left[K(L_1+L_2)\right]}
\\
& = \cos (q_1 L_1) \cos (q_2 L_2) -\frac{s_1^2+s_2^2}{2 s_1 s_2} \sin (q_1 L_1) \sin (q_2L_2), \nonumber
\end{align}
where $K$ is in-plane momentum in the direction of grating and $q_\alpha =\omega/s_\alpha$ for zero damping rate, $\gamma=0$ (see Eq.~\eqref{Eq-wv}). 
Solutions of this equation for $K=0$ are of particular interest, because, they could be excited by normal incident light with zero in-plane momentum. It turns out, however, that only half of frequencies found from Eq.~\eqref{Eq-natural_frequency} show up
in the excitation spectrum provided that DGGS structure has an inversion center. Hence, there are so-called ``bright'' and ``dark'' modes (see analytical description in Refs.~\cite{Gorbenko2024, Khisameeva2025} and numerical analysis in Refs.~\cite{Popov2015, Fateev2019a, Fateev2019b}).

\begin{figure}[h!]
\centering
\includegraphics[width=8.6 cm]{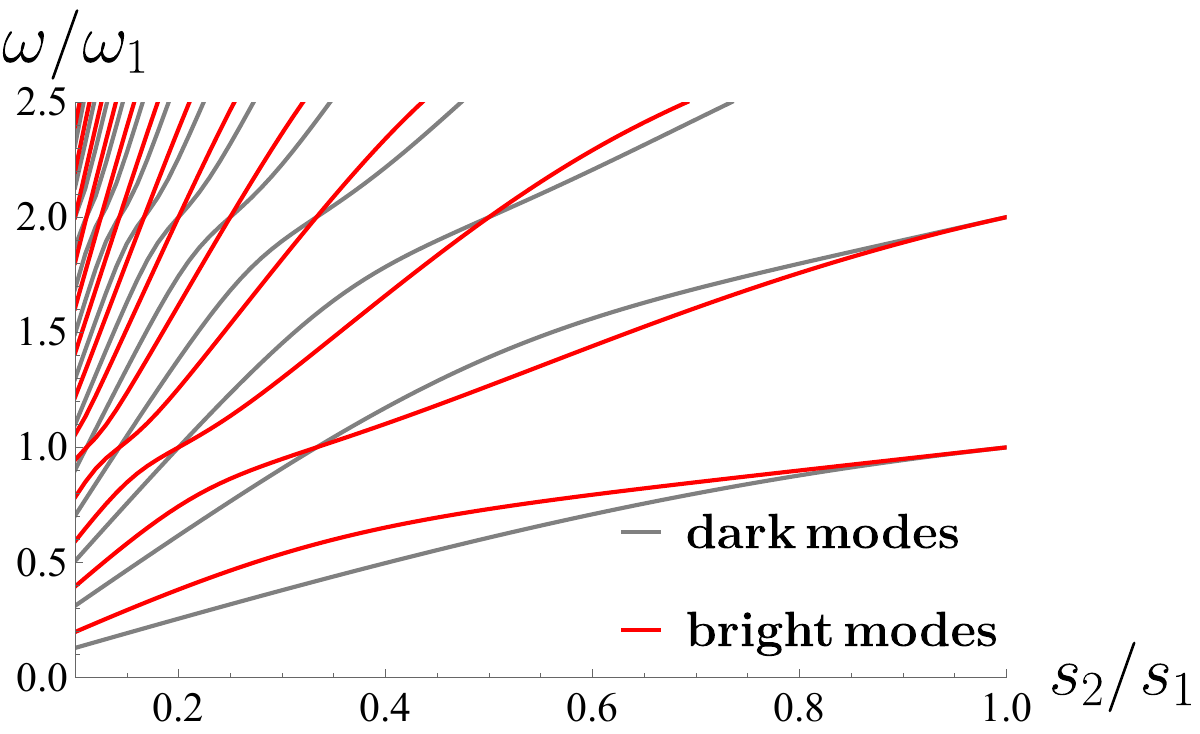}
\includegraphics[width=8.6 cm]{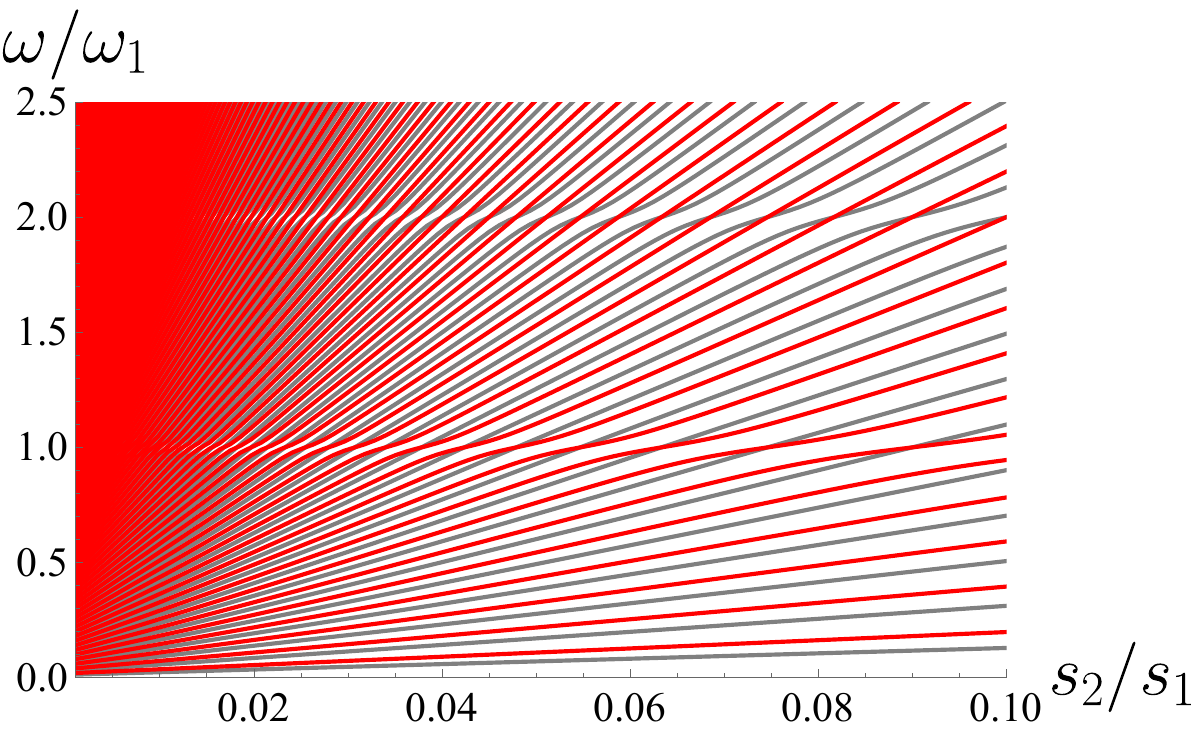}
\caption{(updated from Ref.~\cite{Gorbenko2024}) Dependence of the frequencies $\omega_n^{\rm bright}$ (red lines) and $\omega_n^{\rm dark}$ (grey lines) in the units of $\omega_1=2\pi s_1/L$ on the ratio $s_2/s_1$ for $L_1 = L_2.$ Panel (a) shows interval $0.1<s_2/s_1<1,$ Panel (b) shows interval $0<s_2/s_1<0.1.$ As seen from the panel (b), spectrum of resonances becomes infinitely dense in the limit $s_2 \to 0$ because distance between neighboring levels turns to zero: $\Delta \omega \sim s_2/L_2 \to 0 $. 
}
\label{Fig-bright-dark}
\end{figure}

\begin{figure}[h!]
\centering
\includegraphics[width=6.6 cm]{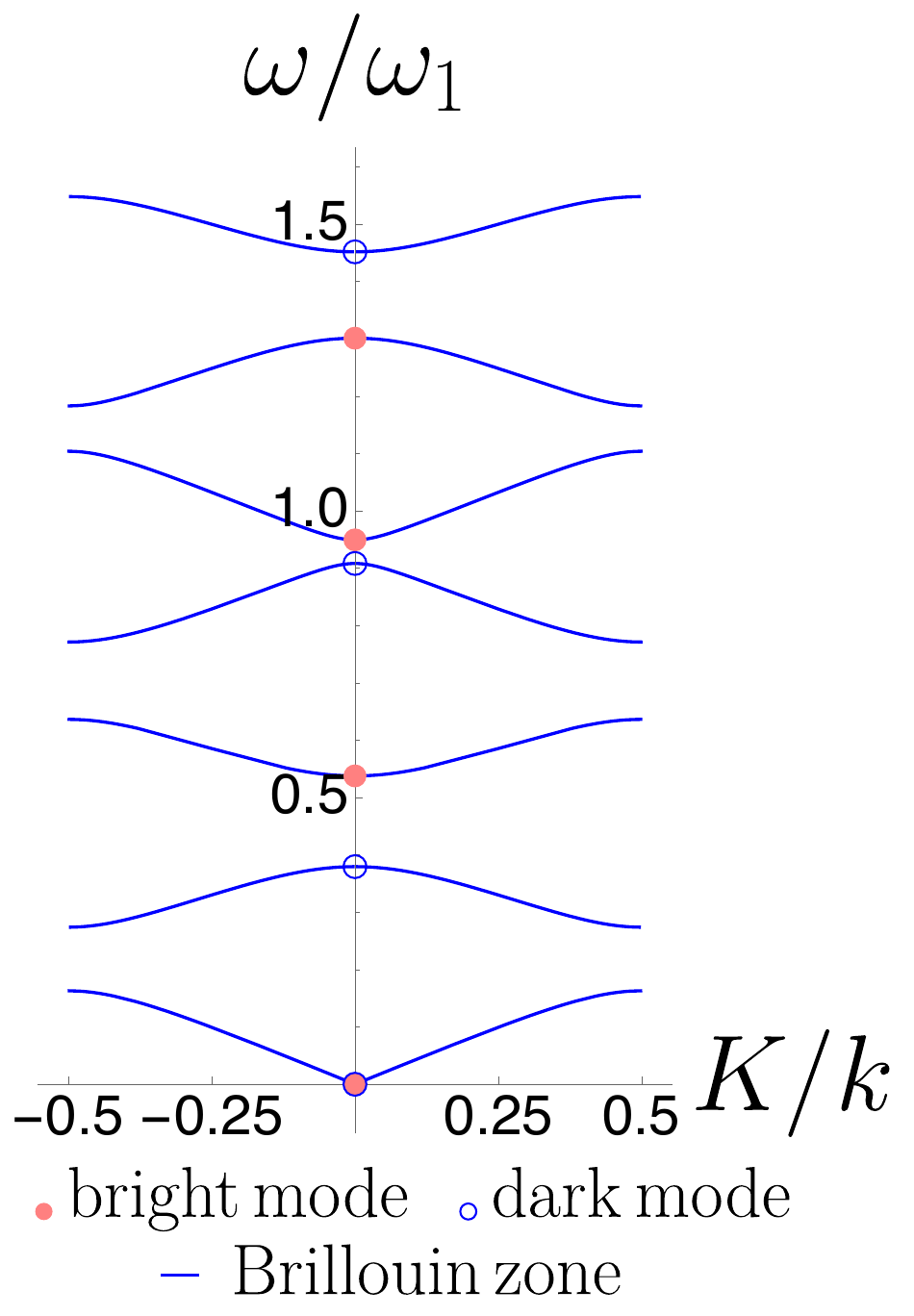}
\caption{(updated from Ref.~\cite{Gorbenko2024}) 
Spectrum of the ideal (i.e. with $\gamma=0$) LPC for $s_2 = 0.3~ s_1, ~L_1=L_2.$ Bright and dark modes are shown at $K=0$ by thick red points and open circles, respectively.
}
\label{Fig-spectrum}
\end{figure}

Let us discuss this issue in more detail. For $K=0$ solutions of Eq.~\eqref{Eq-natural_frequency} can be separated into two types, obeying respectively equations 
\be
\Sigma_{\rm b}=0 \quad \text{or} \quad \Sigma_{\rm d}=0,
\label{sigma_bd_0}
\ee
where 
\begin{align}
 & \Sigma_{\rm b} (\omega) = s_1 \cot{q_1 L_1/2}+s_2 \cot{q_2 L_2/2},
 \label{sigma_b}
 \\
 &\Sigma_{\rm d}(\omega) = s_2 \cot{q_1 L_1/2}+s_1 \cot{q_2 L_2/2},
 \label{sigma_d}
 \end{align}
and one should take $\gamma=0,$ while solving Eq.~\eqref{sigma_bd_0}.
Solutions obeying $\Sigma_{\rm b}=0$ are bright modes that can be excited even for symmetrical modulation, while solutions of $\Sigma_{\rm b}=0$ arise only for asymmetrical modulation without inversion center. As we demonstrate below in the ratchet problem, where asymmetry of the system is a key ingredient of the non-zero photoresponse, both bright and dark plasmonic modes contribute on equal footing, yielding responses of more or less equal amplitudes. Moreover, we will see that in addition to terms that can be unambiguously interpreted as bright or dark contributions, there appear a contribution to the dc photoresponse that describes interference between these two types of modes as well as contributions describing interference between bright modes, dark modes, $n_{\rm ext}$ and $v_{\rm ext}$ . Before closing this subsection, we notice that solutions of Eqs.~\eqref{sigma_b} and \eqref{sigma_d}, 
 $\omega_n^{\rm bright}$ and $\omega_n^{\rm dark}$ (here $n$ numerates solutions) respectively, depend on the ratio $s_2/s_1$ (see detailed discussion in Ref.~\cite{Gorbenko2024}). We illustrate this dependence in Fig.~\ref{Fig-bright-dark} and also show typical spectrum of plasma excitations in Fig.~\ref{Fig-spectrum}. 

\subsection{Linear response to electromagnetic radiation}
Analytical expression for $A_\alpha$ and $B_\alpha$ for $E_0 \neq 0$ can be found by using transfer matrices found in Ref.~\cite{Gorbenko2024}. The details of calculations are delegated to Appendix \ref{AppHD}, where we also present analytical expressions for $A_\alpha$ and $B_\alpha$ for the simplest case $L_1=L_2=L/2$ (general expressions for arbitrary relation between $L_1$ and $L_2$ are quite cumbersome and we do not present them here). 

Two key points of such calculations deserve a short discussion:
\\
\noindent (i) Inhomogeneous part of external field $E_0 h \cos(k x + \phi)$ is spatially periodic with the period $L_1+L_2.$ Hence, we can search for solution, where coefficients $A_\alpha$ and $B_\alpha$ are the same in all cells of the LPC. 
\\
\noindent (ii) As we explained above, plasmonic oscillations in LPC can be separated onto bright and dark modes. For asymmetric modulation needed for the ratchet effect both types of modes are excited. As a result, coefficients $A_\alpha$ and $B_\alpha$ can be presented as
\aleq{
& A_\alpha= \frac{A^{\rm b}_{\alpha}}{\Sigma_{\rm b}}+ \frac{A^{\rm d}_{\alpha}}{\Sigma_{\rm d}},
\\
&
B_\alpha= \frac{B^{\rm b}_{\alpha}}{\Sigma_{\rm b}}+ \frac{B^{\rm d}_{\alpha}}{\Sigma_{\rm d}},
\label{Eq_AB_bd}
}
where formulas for $A^{\rm b,d}_{1,2} $ and $B^{\rm b,d}_{1,2} $ are presented in Appendix~\ref{App-AB12bd}. For small $\gamma,$ 
$A_\alpha$ and $B_\alpha$ show sharp bright and dark plasmonic resonances. Looking at Eqs.~\eqref{vxy} and \eqref{Eq_nv_ext} we see that in addition to the bright and dark plasmonic resonances one could expect three additional peaks encoded in the $n_\alpha^{\rm ext}, v_\alpha^{\rm ext}$: the Drude peak at zero frequency and two plasmonic resonances at frequencies 
$\omega_1=s_1 k,~\omega_2=s_2 k.$ For weak coupling regime, $s_1 \approx s_2$ the frequencies, $\omega_{1,2}$ are close both to each other and to the frequencies of fundamental bright and dark modes, $\omega_{\rm b} =\omega_1^{\rm bright}$ and $\omega_{\rm d} = \omega_1^{\rm dark},$ respectively. However, with increasing coupling all four frequencies are well distinguishable and obey the following inequality: $ \omega_2<\omega_{\rm d}< \omega_{\rm b}<\omega_1.$ Hence, one could expect four plasmonic resonances in addition to the Drude peak. However, as we will see from direct calculations, the resonances at $\omega_{1,2}$ are absent, so that there are three peaks in the dc response: two plasmonic resonances at $\omega_{\rm b}$ and $ \omega_{\rm d}$ as well as the Drude peak. \color{black}

\subsection{Rectification of ac signal}
% {Second order of linearization $\propto E_0^2$, time averaged.}
To find the rectified concentration and velocity $\delta n_{\alpha}$ and $\delta v_{\alpha}$, we average equations Eq.~\eqref{Eq_Navier_Stokes} and \eqref{Eq_continuity} over time thus obtaining in the second order with respect to $E_0$: 
\begin{align}
&s^2_\alpha \frac{\partial }{\partial x} \left[ \delta v_\alpha+ n_\alpha v_\alpha^* + n_\alpha^* v_\alpha \right] = 0, \label{Eq-n1}
\\
&\frac{\partial (v_\alpha v^*_\alpha)}{\partial x} + \gamma \delta v_\alpha + s_\alpha^2 \frac{\partial \delta n_\alpha }{\partial x} = 0.\label{Eq-v1}
\end{align}
These equations should be solved with the boundary conditions found by averaging of Eqs.~\eqref{Eq-BC} over time at the boundary between ``1'' and ``2'': 
\be
J = \langle J_1^{\rm tot} \rangle_t = \langle J_2^{\rm tot} \rangle_t, \quad \langle \epsilon_1 \rangle_t = \langle \epsilon_2 \rangle_t 
\label{Eq-BC2}
\ee
Simple calculation yields:
  $\langle J^{\rm tot}_\alpha \rangle_t \propto s_\alpha^2 (\delta v_\alpha +n_\alpha v_\alpha^* +n_\alpha^* v_\alpha)$ and $\langle \epsilon_\alpha \rangle_t= s_\alpha^2 (\delta n_\alpha + v_\alpha v_\alpha^*)
.$

Physically, Eq.~\eqref{Eq-n1} is responsible for conservation of the dc current in the channel with account of the rectified ac signal, $\langle n v \rangle_t.$ Its solution 
\be
\delta v_\alpha = -(n_\alpha v_\alpha^* + n_\alpha^* v_\alpha) + V_\alpha, 
\label{Eq-dv}
\ee
is expressed in terms of first order ac amplitudes $n_\alpha,~v_\alpha$ and two constant velocities, $V_1$ and $V_2$, describing current flow in the active and passive regions. These velocities are connected with the dc flow in the channel as \be 
J=e N_1 V_1 = e N_2 V_2, 
\label{Eq-J1=J2}
\ee 
where drift velocities are not independent and, with account of Eq.\eqref{Eq-N0}, obey $ 
 s_1^2 V_1=s_2^2 V_2.$
 
Physical meaning of Eq.~\eqref{Eq-v1} is also transparent. This is local Ohm law with driving force given by the sum of the optical force $\p_x (v_\alpha v^*_\alpha),$ arising due to the rectification of ac signal, and static Coulomb field $s_\alpha^2 \p_x \delta n_\alpha $ caused by static density variation. In the absence of voltage applied to the LPC as a whole (open circuit case) both contributions to the driving force are spatially periodic with the period $L$ and, due to the second boundary condition [see Eq.~\eqref{Eq-BC2}] the total force does not have singularities on the boundaries between active and passive regions. Then, integrating Eq.~\eqref{Eq-v1} over cell of the LPC, we get 
\be
\int_0^{L_1}dx~ \delta v_1 + \int_{L_1}^{L_1+L_2} dx~ \delta v_2=0.
\label{Eq-int-dv}
\ee
Substituting Eq.~\eqref{Eq-dv} in Eq.~\eqref{Eq-int-dv} and using Eq.~\eqref{Eq-J1=J2}, we find rectified dc current in the channel: 
\begin{equation}
\begin{aligned}
&J = e \left( \frac{L_1}{N_1}+\frac{L_2}{N_2} \right) ^{-1}
%\nonumber
\\
& \times \left[ \int_0^{L_1}{\! \! \! \! \! \! \left[ n_1 v_1^* + n_1^* v_1 \right] \, dx} + \int_{L_1}^{L_1+L_2}{\! \! \! \! \! \! \! \! \! \! \! \! \! \! \left[ n_2 v_2^* + n_2^* v_2 \right] \, dx} \right].
\end{aligned}
\label{Eq-Jdc}
\end{equation}
% Equation \eqref{Eq-Jdc} can be generalized for arbitrary number of the regions \cite{text1}.

 \subsection{Different contributions to the dc current}

Substituting analytical expressions for $A_{1,2}^{\rm b,d}$ from Appendix \ref{App-AB12bd} into Eq.~\eqref{Eq_AB_bd} 
and using Eqs.~\eqref{vxy},\eqref{Eq_nv_ext} and ~\eqref{Eq-Jdc}, one can find that dc current can be presented as a sum of four terms (see discussion in Appendix \ref{App-relevant}):
\be
J = j_0 + j_{\rm b} +j_{\rm d} + j_{
\rm bd} + {\rm c.c.},
\label{Eq_J}
\ee
where complex currents $j_{\rm b},$ $j_{\rm d},$ and $j_{
\rm bd}$ show singularities at $\omega=\omega_n^{\rm bright}$ and $\omega=\omega_n^{\rm dark}$ for $\gamma=0$ (i.e. resonant peaks at non-zero $\gamma$): 
\be
\begin{aligned}
 & j_{\rm d} \propto \frac{1}{\Sigma_{\rm d} }, \quad
 j_{\rm b} \propto \frac{1}{\Sigma_{\rm b}}, \quad
 j_{\rm bd} \propto \frac{1}{\Sigma_{\rm b} \Sigma_{\rm d}^* }.
 \end{aligned}
 \ee
Analytical expressions for $j_0, j_{\rm b},~j_{\rm d},~j_{\rm bd}$ are presented in Appendix~\ref{App-B-current}. 

Next, in the following sections we will discuss the limiting cases of the Eq.~\eqref{Eq_J}: the low-frequency Drude peak Sec.~\ref{sec-drude}, weak coupling regime, Sec.~\ref{sec-weak}, and strong coupling regime, Sec.~\ref{sec-strong}. In the Section \ref{sec-control}, we will focus on the control of the current by gates, and finally, summarize our results in Section \ref{sec-summary}. 

\section{ Drude peak and weak coupling regime} \label{sec-results}

\subsection{Drude peak} \label{sec-drude}
For small $\gamma$ ($\gamma \ll \omega_{1,2} ~ \gamma \ll \omega_{\rm b,d}$), in the low-frequency limit, $\omega \sim \gamma,$ we get from Eq.~\eqref{Eq_J} the Drude peak 
%\color{red}
\be
J_{\rm Dr}= J_0 \left(\frac{s_1^2 -s_2^2}{s_1^2+s_2^2} \right)\frac{\gamma^2}{\omega^2 + \gamma^2},
\label{Jdrude}
\ee
\be
J_0= -\frac{C e E_0^2 h L \cos{\phi}}{\pi^2 \gamma m}.
\label{J0}
\ee
Its noteworthy that the maximal value of the Drude peak, 
\be
J_{\rm Dr}^{\rm max}=
J_0 \left(\frac{s_1^2 -s_2^2}{s_1^2+s_2^2} \right),
\ee
 tends to zero in the homogeneous case $s_1=s_2$ by contrast to the Drude peak in the transmission coeffcient \cite{Gorbenko2024}. 
Using Eq.~\eqref{Eq-asym_PC}, we find $J_0 ({s_1^2 -s_2^2})/ ( s_1^2+ s_2^2) = \Xi_{\rm PC} \left(-C e L /\gamma m \pi^2 \right),$ so that the current is proportional to the asymmetry parameter: $J \propto \Xi_{\rm PC}.$

Equation ~\eqref{J0}, deserves a special comment. The current $J_0$ does not depend on concentration. 
On the other hand, in the limit $N_2 \to 0$ region 2 has infinite resistance and can not conduct dc current. 
A similar problem arises during optical excitation of single field-effect transistors \cite{Gutin2012}.
To resolve the contradiction, one should recall that Eq.~\eqref{J0} is derived from perturbation theory, which assumes that the oscillating concentration is smaller than the average, i.e., $n_{1,2} \ll 1$. 
It is not difficult to show that for fixed values of the field $E_0$ and velosity $s_1$, perturbation theory breaks down at sufficiently small $s_2$.
In particular, it is straightforward to show that the condition $n_2 \ll 1$ yields 
\be
\frac{e E_0}{m (i \omega +\gamma) }
\frac{s_1}{s_2^2} \ll 1.
\label{ineq1}
\ee
If $s_2$ becomes so small that inequality \eqref{ineq1} is not satisfied, it is necessary to use a non-perturbative approach, similar to the one used in \cite{Gutin2012}. The corresponding calculation will be presented elsewhere.

\subsection{Weak coupling, $\Delta \ll 1$} 
\label{sec-weak}

Next, we assume that the 
electron concentration is weakly modulated, $(N_1 - N_2) \ll N_1,$ and, consequently, $\Delta \ll 1.$ 
In this case, $\omega_{\rm b} \approx \omega_{\rm d} \approx \omega_1 \approx \omega_2  .$ 
Having in mind that in the homogeneous case, when $\Delta\equiv 0,$ all these frequencies   coincide, it is convenient to change notation $\omega_1 \to \omega_0,$  
where   frequency  
\be
\omega_0 = k s_1 = 2 \pi s_1/L
\ee
has a physical meaning of plasmonic resonance frequency in the homogeneous system.
Non-zero ratchet current appears already within the first order with respect to $\Delta.$ Direct expansion of Eq.~\eqref{Eq_J} over $\Delta$ yields
\be
J_{\rm weak}^0 = J_0 \Delta
\frac{\gamma^2}{\omega^2+\gamma^2} \frac{\omega_0^4}{(\omega^2 - \omega_0^2)^2 + \gamma^2 \omega^2}.
\label{Eq_Jweak}
\ee
Having in mind that for weak coupling, $ (s_1^2- s_2^2)/ (s_1^2+ s_2^2) \approx \Delta ,$ we reproduce for small $\gamma$ the Drude peak, Eq.~\eqref{Jdrude}, at low frequencies $\omega \sim \gamma$. Equation \eqref{Eq_Jweak} also shows 
a fundamental plasmonic resonance at $\omega = \omega_0$. The amplitudes of two peaks are connected as 
\be
\frac{J_{\rm weak}^0 (\omega = \omega_0)}{J_{\rm Dr}^{\rm max} (s_2 \approx s_1) } = \frac{\omega_0^2}{\gamma^2+\omega_0^2}
.
\label{Eq-Jweak_ratio}
\ee
This ratio approaches $1$ for resonant case, $\omega_0 \gg \gamma$ and tends to zero for nonresonant case, $\omega_0 \ll \gamma$. In the resonant case, expression for the fundamental peak becomes
\be
J_{\rm weak}^{\rm pl} \approx \frac{J_0 \Delta (\gamma/2)^2}{ (\omega-\omega_0)^2 + (\gamma/2)^2}.
\label{J_weak_pl}
\ee
It is worth noting that the width of the Drude peak, $\gamma,$ is twice larger than the width of the plasmonic peak, $\gamma/2.$ We also notice, that Eq.~\eqref{Eq_Jweak} does not contain any information about bright and dark modes and coincides (after correct redefinition of $J_0$) with the equation obtained earlier by using perturbative approach within the minimal model \cite{Rozhansky2015}.

\begin{figure}[h!]
\includegraphics[width=0.5 \textwidth]{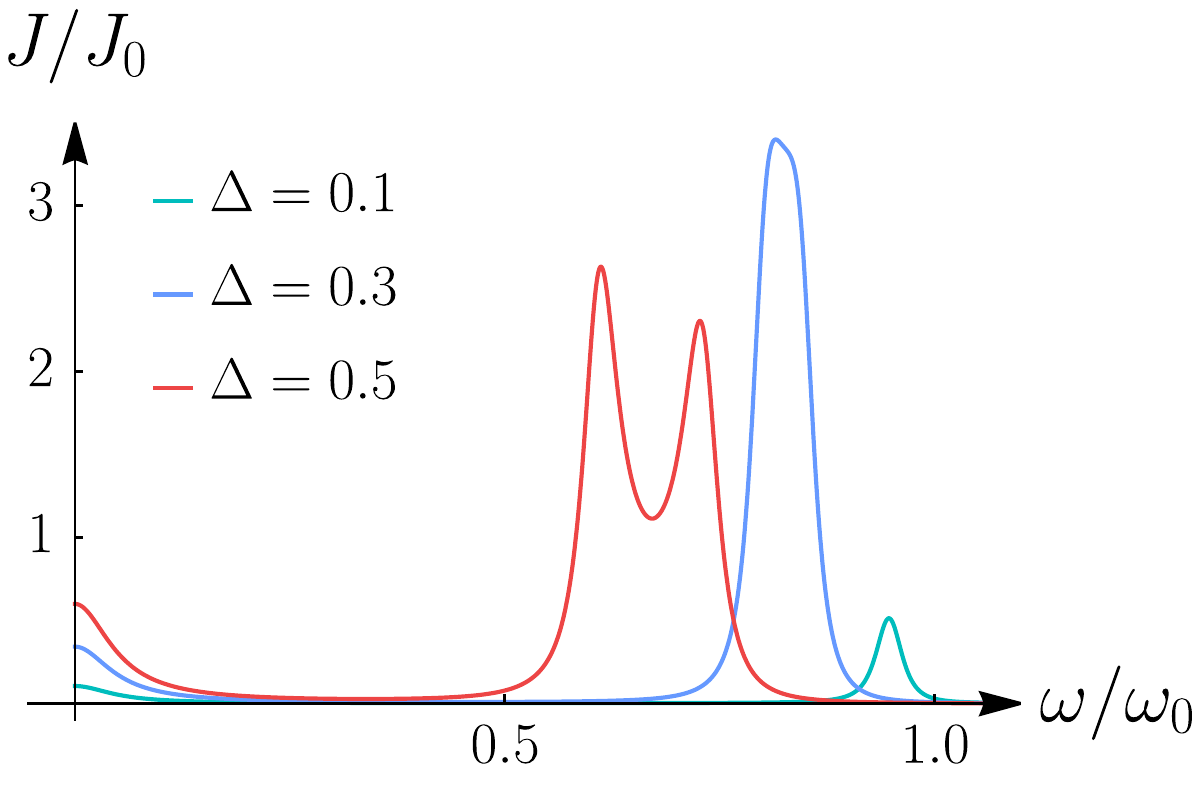}
\caption{ 
Plot of dc current, $J$ on frequency $\omega/\omega_0$ for the resonant case $\gamma/\omega_0 = 0.05 $ and different values of $\Delta$: $\Delta=0.1,~ \Delta=0.3$ and $\Delta=0.5$.}
\label{fig_weak}
\end{figure}

Let us now discuss most interesting non-perturbative regime. 
In Fig.~\ref{fig_weak} we plotted $J$ calculated with the use of the exact formula \eqref{Eq_J} within the frequency range including the Drude peak and the fundamental frequency $\omega_0$ for different values of $\Delta.$ 
One can see that significant deviations from the approximate formula \eqref{Eq_Jweak} arise even at small $\Delta$ ($\Delta \sim 0.1 $), whereby only the plasmon peak is strongly modified, while the Drude peak remains practically unchanged.

Let us discuss this point in more detail focusing on the modifications of the fundamental plasmonic resonance. From the Fig.~\ref{fig_weak} we see three effects of increasing $\Delta $: 
\begin{itemize}
 \item resonant frequency decreases below the value $\omega_0$; 
 \item the amplitude of the plasmonic peak becomes essentially greater than the amplitude of the Drude peak already at sufficiently small $\Delta$ in the evident contradiction with Eq.~\eqref{Eq-Jweak_ratio};
 \item with increasing $\Delta,$ plasmonic resonance splits onto two peaks. 
\end{itemize}

In order to get more accurate approximate formula for $J_{\rm weak}^{\rm pl},$ which captures properties mentioned above, we have to go beyond linear-in-$\Delta$ approximation. We expect that 
spectrum of the dc current $J$ would demonstrate resonances both at $\omega=\omega_n^{\rm bright}$ and $\omega=\omega_n^{\rm dark}$. Let us focus on the fundamental modes $\omega_{\rm b}=\omega_1^{\rm bright}$ and $\omega_{\rm d}=\omega_1^{\rm dark}.$ 

Solving approximately Eqs.~\eqref{sigma_bd_0} at small $\Delta,$ we get
\be
\omega_{\rm b} = \omega_0 \left[ 1 - \frac{\Delta}{2}\right] + O(\Delta^4) ,
\label{Eq_wb_split}
\ee
\be
\omega_{\rm d} =\omega_0 \left[ 1 - \frac{\Delta}{2} -\frac{\Delta^2}{2} - \frac{\Delta^3}{4}\right] + O(\Delta^4) .
\label{Eq_wd_split}
\ee
Hence, in the weak coupling regime, both frequencies 
 are close to  $\omega_0$ and difference between $\omega_{\rm b}$ and $\omega_{\rm d}$ appears only in the second order with respect to $\Delta.$ 
Expansions of $\Sigma_{\rm b,d}$ close to these frequencies read 
\be
\Sigma_b \approx -\frac{L}{2} \left[\left(\omega-\omega_b\right) + i \frac{\gamma}{2} \right] , \, \, \Sigma_d \approx -\frac{L}{2} \left[\left(\omega-\omega_d\right) + i \frac{\gamma}{2} \right]
\ee
(here we neglect linear-in-$\Delta$ correction to imaginary parts of $\Sigma_{\rm b,d}$).
Substituting these expansions into Eq.~\eqref{Eq_J}, after some algebra we obtain a more correct asymptotical equation:
\be
J_{\rm weak}^{\rm pl} = J_0 \Delta \frac{ (\gamma^2/4) \left[ (\omega-\omega_0)^2+\gamma^2/4 \right]}{ \left[(\omega-\omega_b)^2+\gamma^2/4 \right] \left[(\omega-\omega_d)^2+\gamma^2/4\right]},
\label{J_weak_pl_1}
\ee
which coincides with Eq.~\eqref{J_weak_pl} for $\omega_{\rm b,d} \to \omega_0.$

From Eq.~\eqref{J_weak_pl_1}, we find that the maximum value of current is reached at frequencies $\omega_{\rm b,d}$ 
\be
J^{\rm max
} \approx 
J_{\rm weak}^{\rm pl}( \omega_b) \approx J_{\rm weak}^{\rm pl} (\omega_d) \approx J_0 \frac{\Delta \left(\Delta^2 + \gamma^2/\omega_0^2 \right)}{ \Delta^4 + \gamma^2/\omega_0^2 }
\label{Eq_J_weak_res}
\ee
The ratio of this current to the amplitude of the Drude peak is given by 
\be
\frac{J^{\rm max}}{J_{\rm Dr}^{\rm max}(s_1 \approx s_2)}
= \frac{1+ \frac{\omega_0^2}{\gamma^2 } \Delta^2 }{1+ \frac{\omega_0^2}{\gamma^2 } \Delta^4}
\label{Jm_JD}
\ee

The latter two equations deserve some discussion. With increasing $\Delta$ (for fixed $\gamma/\omega_0 \ll 1$), the current $J^{\rm max
}$ is proportional to $\Delta$ for $\Delta \ll \gamma/\omega_0, $ then starts to increase much faster, $\propto \Delta^3$ , for $ \gamma/\omega_0 \ll \Delta \ll \sqrt{\gamma/\omega_0}, $ rich maximum at $\Delta \approx 3^{1/4} \sqrt{\gamma/\omega_0} $ and, next, decays as $\propto 1/\Delta$ at $ \Delta \gg \sqrt{\gamma/\omega_0}, $ when the plasmonic resonance splits into bright and dark peaks. 

\begin{figure}[h!]
\hspace*{-0.5cm}
\includegraphics[width=0.48\textwidth]{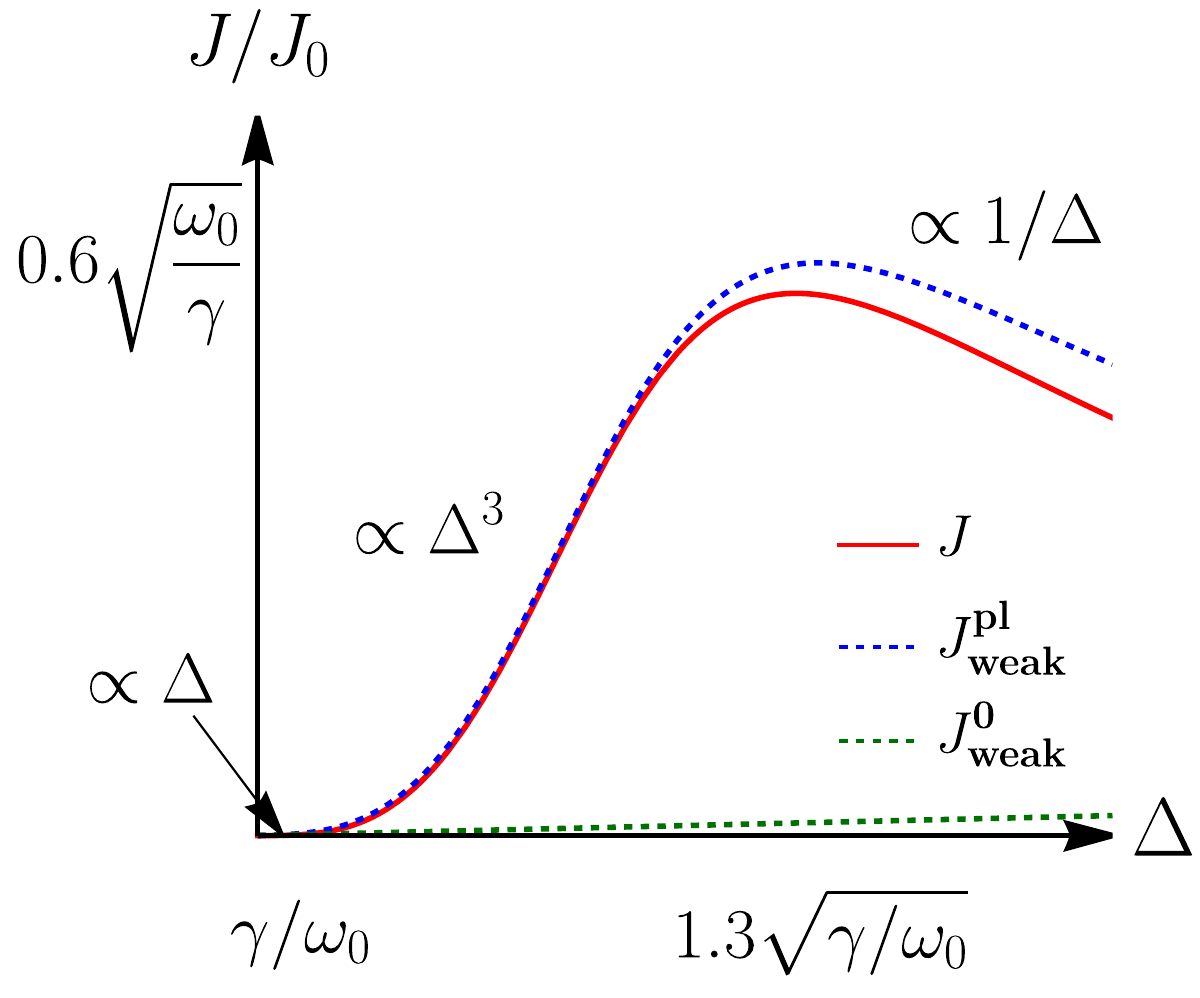}
\caption{ 
Plot of exact and approximate expressions for dc current at resonant frequency: $J(\omega = \omega_b)$ (red curve, $\omega_b$ is given by Eq.\eqref{Eq_wb_split} and $J$ by Eq.\eqref{Eq_J}), $J_{\rm weak}^{\rm max}$ (dashed blue curve, see Eq.~\eqref{Eq_J_weak_res}) and linear-in-$\Delta$ approximation $J_{\rm weak}^0$ (dashed green curve, see Eq.~\eqref{Eq_Jweak}). Plots are made for the resonant case $\gamma/\omega_0 = 0.01$. To simplify plot we replaced exact numerical values with approximations: $3^{3/4}/4 \approx 0.6$, $3^{1/4} \approx 1.3$.}
\label{fig_weak_amplitude}
\end{figure}

Equation \eqref{Eq_J_weak_res} is plotted in Fig.\ref{fig_weak_amplitude}. 
Importantly, there is a parametrically large interval,
$ \gamma/\omega_0 \ll \Delta \ll \sqrt{ \gamma/\omega_0}, $ where bright and dark peaks overlap and form a single peak at 
$\omega_{\rm b} \approx \omega_{\rm d} \approx \omega_0(1-\Delta/2)$, which
rapidly ($\propto \Delta^3$) increases with $\Delta.$ Within this interval Eq.~\eqref{J_weak_pl_1} simplifies
\be
J_{\rm weak} ^{\rm pl} =J_0 \Delta \frac{(\gamma^2/4) \left[(\omega - \omega_0)^2+\gamma^2/4\right]}{ \left[(\omega - \omega_0[1-\Delta/2])^2+\gamma^2/4 \right]^2} .
\label{Eq_Jweakweak}
\ee
The frequency shift of the resonance is given by $-\omega_0 \Delta/2$.

\begin{figure}[h!]
\hspace*{-0.5cm}
\includegraphics[width=0.48\textwidth]{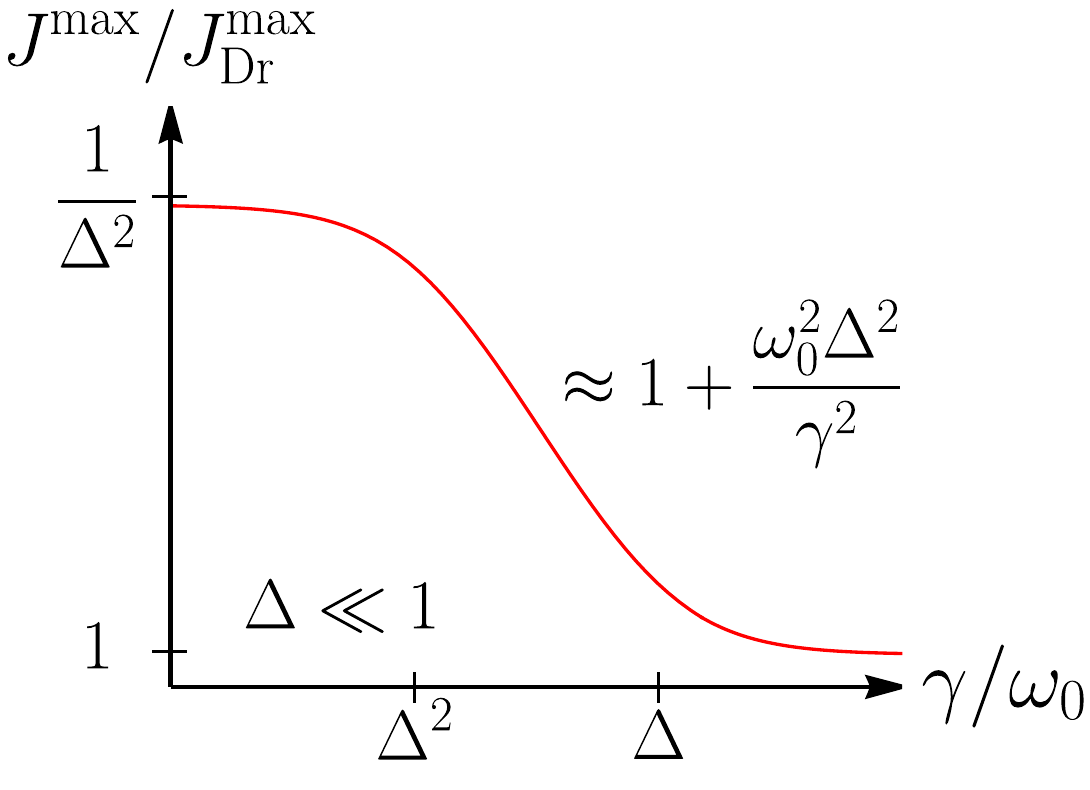}
\caption{ 
Schematic plot of the ratio of the amplitude at the resonant frequency to the Drude peak on $\gamma/\omega_0$ for weak coupling, $\Delta \ll 1.$ For small $\gamma/\omega_0<\Delta^2$ bright and dark modes are distinguishable. For $\Delta^2 < \gamma/\omega_0$ dark and bright modes are merged into single plasmonic resonance. (Here we limit ourselves with regime $\gamma \ll \omega
_0$.)  }
\label{fig_weak_amplitude_gamma}
\end{figure}

The ratio of plasmonic and Drude peak amplitudes in this range of $\Delta$ can be found from Eq.~\eqref{Jm_JD} where one can neglect term $\omega_0^2\Delta^4/\gamma^2$ in the denominator: 
\be
\frac{J_{\rm weak}^{\rm pl} [\omega = \omega_0(1-\Delta/2 )]}{J_{\rm Drude}^{\rm max}} = 1 + \frac{\omega_0^2}{\gamma^2} \Delta^2.
\label{Eq-Jweakweak_ratio}
\ee
Giant increase of the plasmonic peak  is illustrated in Fig.~\ref{fig_weak} and Fig.~\ref{fig_weak_amplitude_gamma}.
 
 Analyzing different contributions to the current entering Eq.~\eqref{Eq_J},  one can see that the main contribution to $J$ comes from the last two terms, $j_{\rm bd}$ (interference of dark and bright plasmons) and $j_{\rm b}$ (interference of  bright plasmons with $n^{\rm ext}$ and $v^{\rm ext}$).
 Hence, interference of bright and dark modes leads to giant enhancement of the plasmonic response.

\begin{figure}[h!]
\includegraphics[width=0.5 \textwidth]{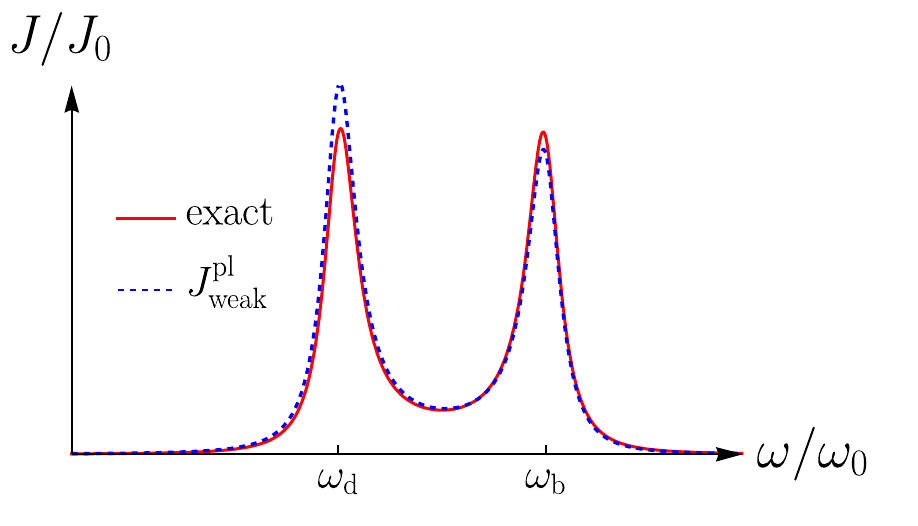}
\caption{ 
Plot of exact expression for current, Eq.~\eqref{Eq_J} (red solid line) and resonant approximations of dark and bright fundamental mode splitting with Eq.~\eqref{J_weak_pl_1}
 at $ \Delta \gg \sqrt{\gamma/\omega_0} $ (figure is plotted for $\gamma = 0.001 \omega_1$ and $\Delta = 0.1$).}
\label{fig_weakBD}
\end{figure}

To close this section, we finally note that the current in a cleaner structure is also well described by approximate Eq.~\eqref {J_weak_pl_1} for not too large $\Delta$ as illustrated in Fig.~\ref{fig_weakBD} for $\Delta =0.1$ With increasing $\Delta$, the split bright and dark peaks become slightly different in amplitude (see Fig.~\ref{fig_weak}) which is not described by simplified Eq.~\eqref {J_weak_pl_1}.

\begin{figure}[h!]
\includegraphics[width=0.5 \textwidth]{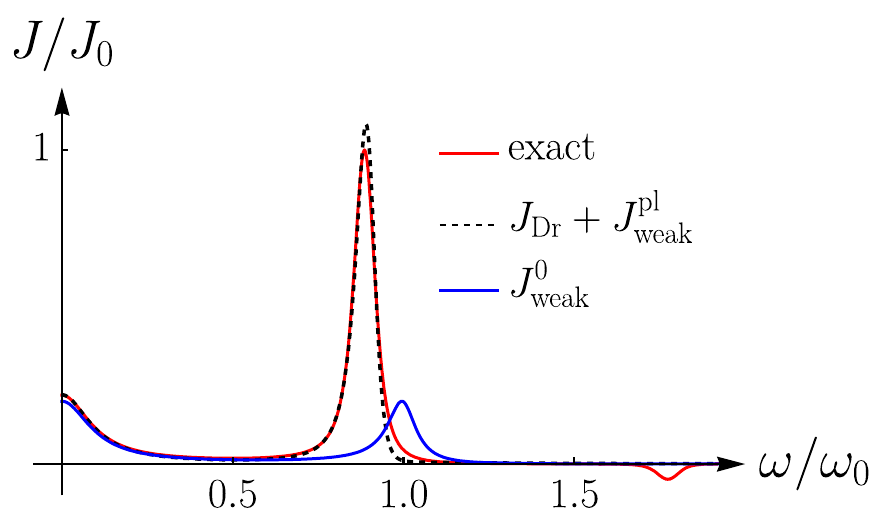}
\caption{ 
Plot of exact expression for current, Eq.~\eqref{Eq_J} (red solid line) together with approximate Eq.~\eqref{Eq_Jweakweak} (dashed black line), which illustrates giant enhancement of the plasmonic peak for $ \gamma/\omega_0 \ll \Delta \ll \sqrt{\gamma/\omega_0} $ (figure is plotted for $\gamma = 0.1~ \omega_1$ and $\Delta = 0.2$). 
One can also see the second harmonic contribution at $\omega \approx 1.75 \omega_0,$ which is negative and rather weak.  
}
\label{fig_weakweak}
\end{figure}

\section{ Strong coupling regime, $X=1-\Delta \ll 1$} 
\label{sec-strong} 
 In this section, we discuss strong coupling regime, when, $ \Delta \to 1, $ so that 
$$X=1-\Delta \ll 1.$$
For convenience, we   use below dimensionless variables:
\be
\Omega = \frac{\omega}{\omega_0},~\Gamma=\frac{\gamma}{\omega_0} \ll 1. \ee

\begin{figure}[h!]
\includegraphics[width=0.4 \textwidth]{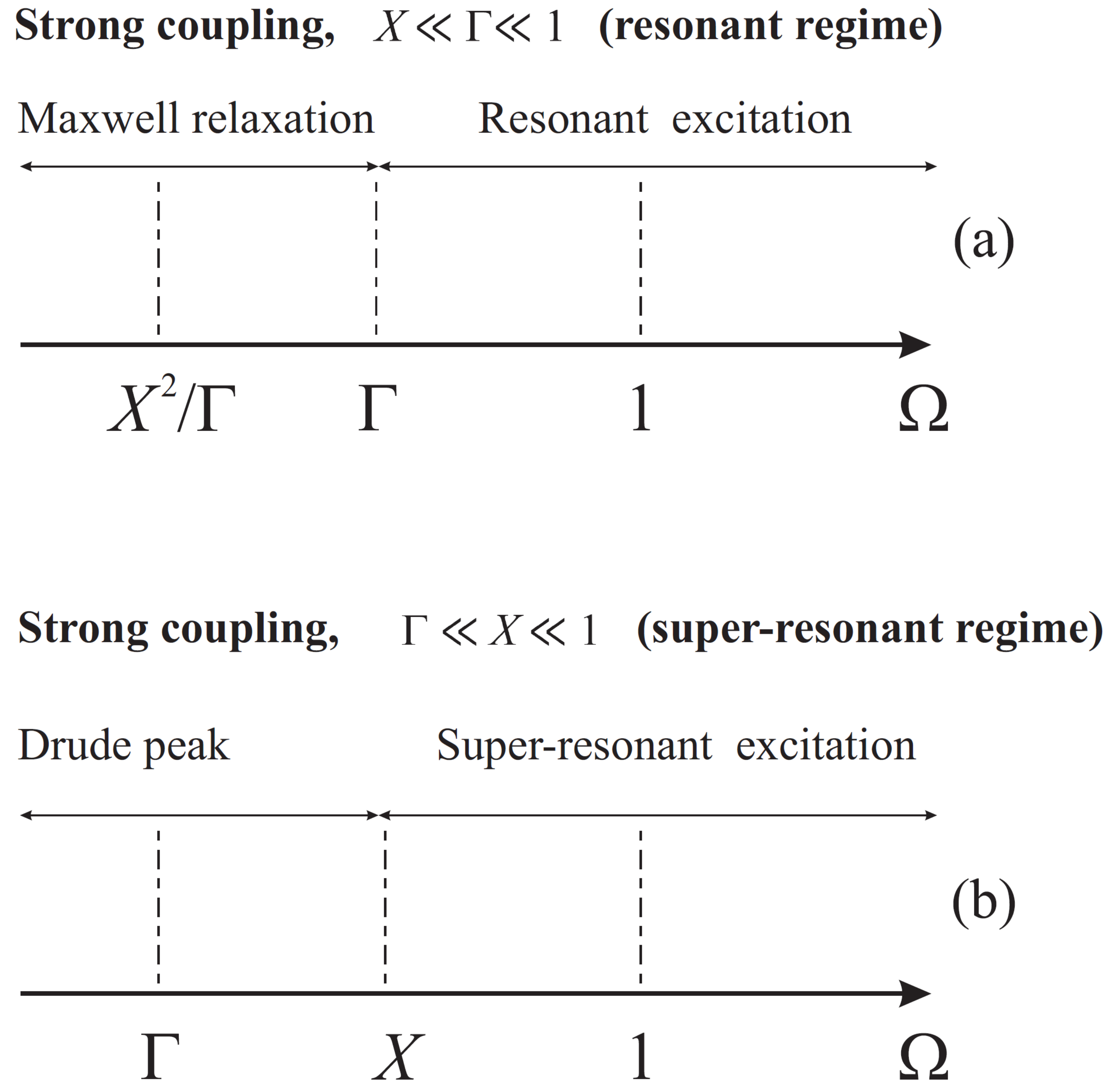}
\caption{Characteristic frequency scales for resonant (a) and super-resonant (b) regimes. }
\label{fig_diagram_strong}
\end{figure}

The results of further calculation depends on relation between $X$ and $\Gamma$ (i.~e. between $\omega_2$ and $\gamma$). For $\Gamma \gg X,$ resonances with the frequency $\omega_2$ and its harmonics overlap, so that dc response can only show resonances related to $\omega_1$ and its harmonics (for $\Omega \gtrsim \Gamma $). Following Ref.~\cite{Gorbenko2024}, we call this regime resonant. In the opposite case, $\Gamma \ll X,$ resonances related to the frequency $\omega_2$ becomes visible in the dc current. This regime is called ``super-resonant'' \cite{Gorbenko2024}. Fig.~\ref{fig_diagram_strong} shows characteristic scales corresponding to resonant (Fig.~\ref{fig_diagram_strong}a) and super-resonant (Fig.~\ref{fig_diagram_strong}b) regimes. 
Next, we discuss both regimes in detail. 

\subsection{Resonant regime, $ X\ll \Gamma \ll 1 $}
In this case, one can send in all exact formulas (see Eq.\eqref{Eq_J} and equations in Appendixes \ref{App-AB12bd} and \ref{App-B-current}) $\cot(q_1 L/4) \to - i$ and $\cot(q_1^* L/4) \to i.$ In thus obtained equation, one can send $X \to 0,$ finally arriving at
\be
\begin{aligned}
\frac{J_{\rm strong}}{J_0}& =\frac{\Omega+ \sqrt{\Omega^2+ \Gamma^2}}{4 (\Omega+ i \Gamma) (1-\Omega^2 + i \Omega \Gamma)} 
\\
& \times \frac
{\cot\left({\pi \sqrt{\Omega(\Omega - i \Gamma)}}/{2}\right)}{\cot\left({\pi \sqrt{\Omega(\Omega +i \Gamma)}}/{2}\right)} ~+~{\rm c.c.} 
\end{aligned}
\label{eq-res}
\ee
This equation contains resonances at frequencies $ \Omega= n $:
\begin{align}
&\frac{J_{\rm strong}^{n=1}}{J_0} = \frac{-2\delta \Omega_1 + \Gamma^2+ \delta\Omega_1^2}{ \Gamma^2+ 4 \delta \Omega_1^2}, \label{eq-1st-res}
 \\
 & \frac{J_{\rm strong}^{n\neq 1}}{J_0} =\frac{1}{n^2-1}\frac{ \Gamma^2 - 4 \delta \Omega_n^2 }{ \Gamma^2+ 4 \delta \Omega_n^2}.
 \label{eq-nth-res}
 \end{align}
Here, $\delta \Omega_n=\Omega-n$ and we assume $\delta\Omega_n \sim \Gamma \ll 1. $
Since the fundamental resonance with $n=1$ is bigger than the ones with $n \neq 1,$ we keep for $n=1$ not only leading term ($-2 \delta \Omega_1$ in the numerator) but also sub-leading terms ($\Gamma^2+\delta\Omega_1^2  $ in the numerator).

\begin{figure}[h!]
\includegraphics[width=0.45
\textwidth]{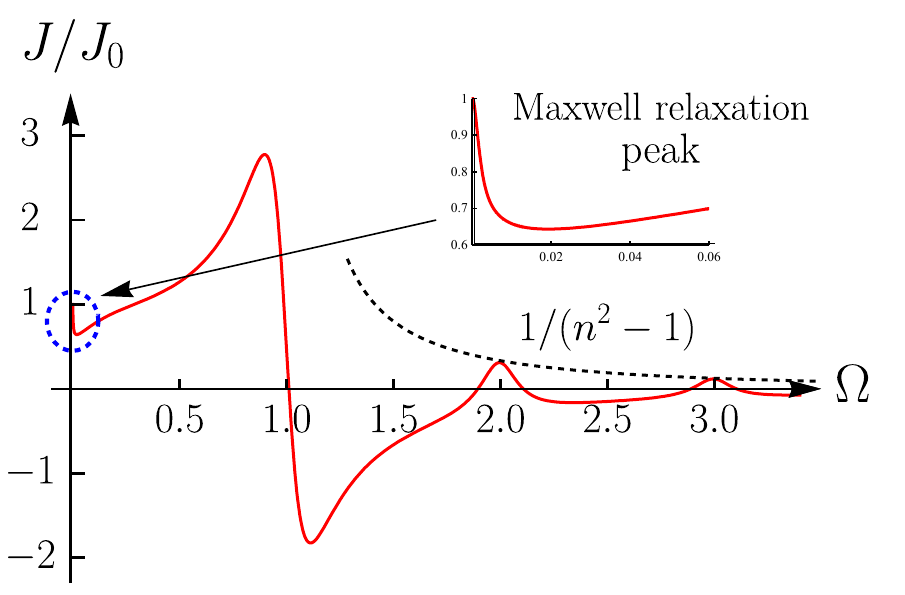}
\caption{Photo-response in the resonant regime for $\Gamma=0.2$ and $X=0.01$. Resonances at high harmonics, $n=2,3,\dots$, decay with $n$ as $1/(n^2 -1)$ and have symmetric form (see Eq.~\eqref{eq-nth-res}) by contrast to resonance at fundamental frequency, $n=1,$ which is parametrically bigger by a factor $\sim 1/\Gamma$ and has asymmetric Fano-like shape (see Eq.~\eqref{eq-1st-res}). At very low frequency there is a sharp peak caused by the Maxwell relaxation (see Eqs.~\eqref{eq-maxwell-1} and \eqref{eq-maxwell-2}). 
}
\label{fig_strong_resonant}
\end{figure}

In the low-frequency limit, $\Omega \sim \Gamma \ll 1$ Eq.~\eqref{eq-res} simplifies
\be
\frac{J_{\rm strong}}{J_0} = \frac{1}{2} \left( 1+\frac{\Omega}{\sqrt{\Omega^2 + \Gamma^2}}\right).
\label{reson1}
\ee

Interestingly, for even lower frequencies, $\Omega \sim X^2 /\Gamma \ll \Gamma$ ($\omega \sim \omega_2^2/\gamma$) there is another regime that is not captured by Eq.~\eqref{eq-res} which was obtained by assuming $X \to 0.$ Physics behind this regime is the Maxwell relaxation in the region 2 with the characteristic rate 
\be
\tau_{\rm M}^{-1} \sim \omega_2^2/\gamma. 
\ee
To get analytical expression in this regime, one can make  the following replacement in general formula \eqref{Eq_J}, $\omega \to \omega t^3,\gamma \to \gamma t,~X \to X t^2 $ and take the limit $t \to 0.$ This yields
\be
\frac{J_{\rm strong}^{\rm M}}{J_0} \approx f(y),
\label{eq-maxwell-1}
\ee
where $y=\sqrt{\omega \tau_{\rm M}}= \sqrt{\Omega \Gamma}/X $ and dimensionless function $f$ reads
\be
\begin{aligned}
f(y)&=\frac{1}{2}+
\frac{\pi y}{4 (1+ 4 y^4)}
\\
&\times \frac{(1-2 y^2) \sin(\pi y) +(1+2 y^2 )\sinh(\pi y)}{\cosh(\pi y) -\cos(\pi y)}.
\end{aligned}
\label{eq-maxwell-2}
\ee

Results of calculations are illustrated in Fig.~\ref{fig_strong_resonant}, where both fundamental peak at $\Omega=1$ and two higher modes with $\Omega=2$ and $\Omega=3$ are shown together with very narrow peak at zero frequency arising due to the Maxwell relaxation in the low-conducting region ``2''. 

\subsection{Super-resonant regime, $ \Gamma\ll X \ll 1 $.}
\begin{figure}[h!]
\includegraphics[width=0.4
\textwidth]{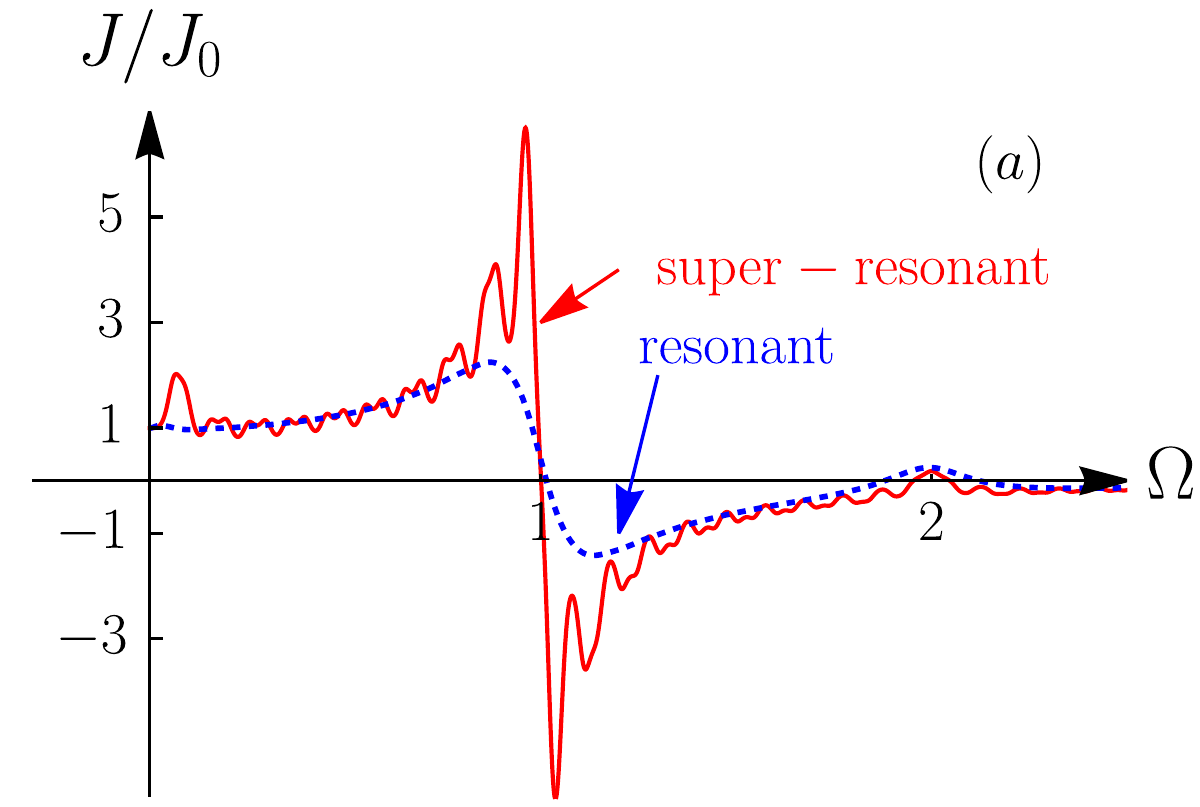}
\includegraphics[width=0.4
\textwidth]{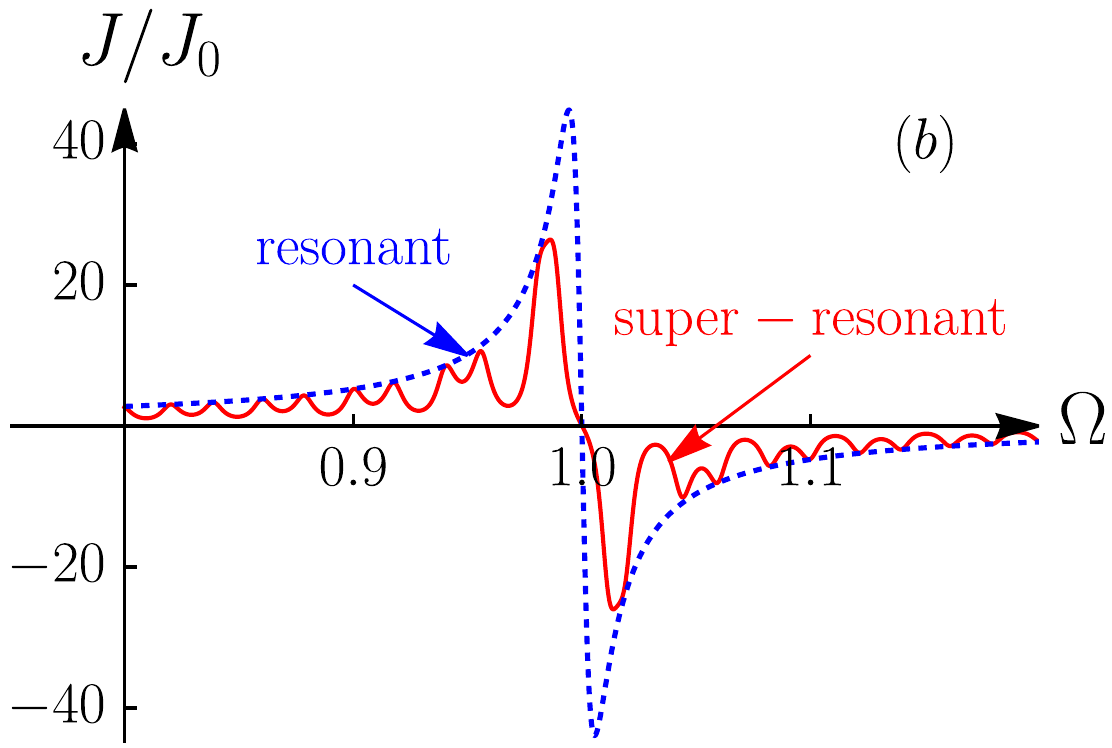}
\includegraphics[width=0.41
\textwidth]{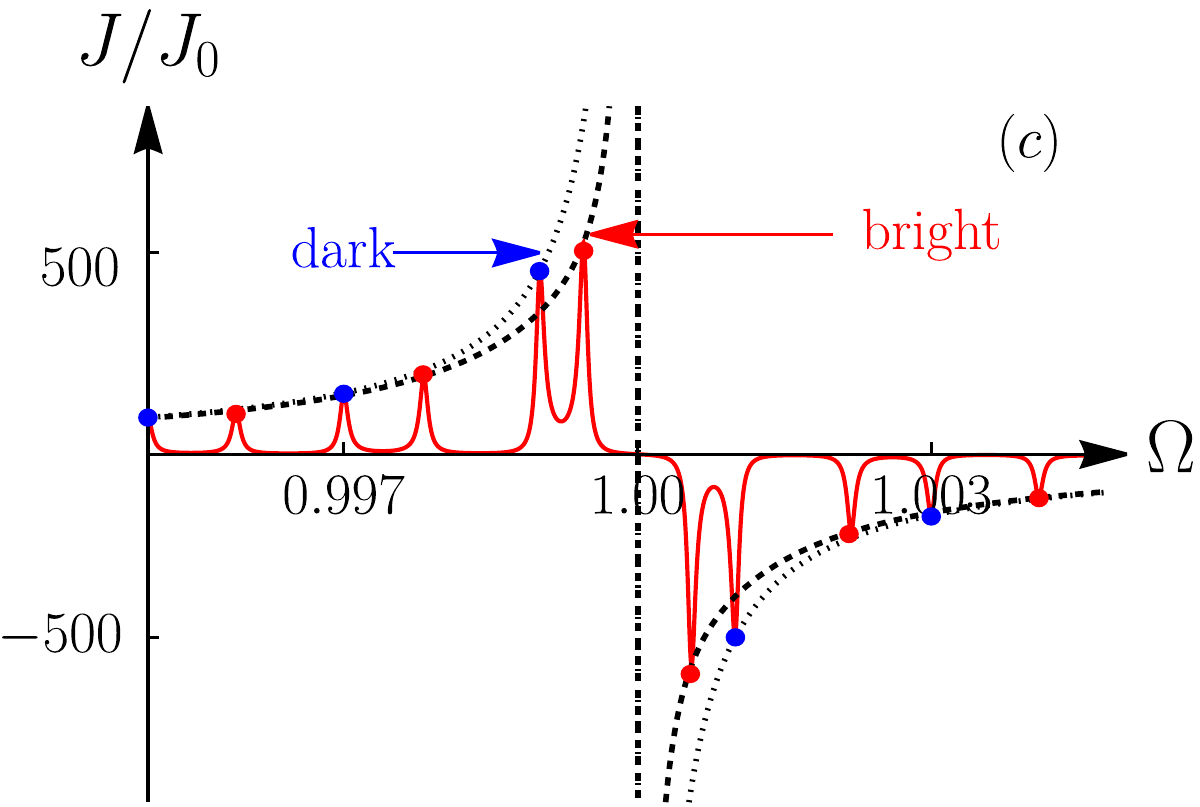}
\caption{
Appearance of fine super-resonant structure with increasing $X/\Gamma$: (a) dashed line -- resonant response at $\Gamma=0.2$ and $X=0.05$, thick line -- super-resonant response at the same $X=0.05$ but smaller damping rate, $\Gamma=0.05$ ; (b) 
splitting of the fundamental resonance with $n=1$ into super-resonances with increasing $X$ for fixed $\Gamma.$ Dashed line -- resonance curve for $\Gamma=0.01$ and $X=0.001,$ thick line -- super-resonant peaks for $\Gamma=0.01$ and $X=0.02$; (c) thick line -- super-resonant response for $\Gamma=10^{-4}$ and $X=10^{-3}$, dashed  and dotted curves -- envelopes of the bright and dark modes, respectively.} \label{fig_strong_super}
\end{figure}

Next, we assume that $\Gamma$ is smaller than $X.$ In this case, fine structure related to lower frequency $\omega_2$ becomes visible in the dc photo-response as illustrated in Fig.~\ref{fig_strong_super} within interval $0<\Omega<3.5$ (Fig.~\ref{fig_strong_super}a) and in more detail near the fundamental resonance with $n=1$ for $0.8<\Omega <1.2$ (Fig.~\ref{fig_strong_super}b). 

Let us describe analytically transition from the resonant regime to the super-resonant one, considering vicinity of the fundamental resonant mode, i.e. assuming that $\delta \Omega_1=\Omega-1 \ll 1.$
To this end, we 
 make the following substitution into Eq.~\eqref{Eq_J}, $\Omega=1- t \delta \Omega_1,~\Gamma \to t \Gamma ,~ X \to t X, $ then consider the limit $t \to 0,$ keeping leading $\propto 1/t$ terms, neglect in the arguments of trigonometric functions small terms $ \propto \Gamma^2/ X $ and, finally, put $t=1,$ thus arriving at \aleq{
&\frac{J_{\rm strong}^{n=1}}{J_0}=-\frac{1}{4}\frac{\sinh \left(\frac{\pi \Gamma }{2 X}\right)}{\sinh \left(\frac{\pi \Gamma }{ 2 X}\right)-i \sin \left(\frac{\pi \Omega}{X}\right)}
 \\
 &\times \frac{1 }{ \left[i \Gamma/2 + \Omega-1 -4 X \cot \left(\frac{\pi (i
 \Gamma +2 \Omega)}{4 X}\right)\right]}+ {\rm c.c.}
 \label{eq-fundam-super}
 }
(here for compactness, we restore notation $\Omega$ instead of $\delta \Omega_1$). Equation~\eqref{eq-fundam-super} is valid for arbitrary relation between $\Gamma$ and $X.$ In the limit $\Gamma \gg X, $ we get $\-2 \delta \Omega_1/(\Gamma^2 + 4\delta \Omega_1^2), $ i.e. restore leading term of Eq.~\eqref{eq-1st-res}. In the opposite limit, $\Gamma \ll X,$ one can expand numerator and denominator of Eq.~\eqref{eq-fundam-super} up to the linear order with respect to $\Gamma,$ and  get formula describing super-resonant oscillations for $ |\Omega-1| \ll 1$:
\be
\frac{J_{\rm sup}^{n=1}}{J_0}=
\frac{1}{4}\frac{ \frac{i \pi^2 \Gamma}{8 X}}{ B (\Omega) D(\Omega) + \frac{i \pi^2 \Gamma}{8 X} C(\Omega) } +c.c.,
\label{eq-super-2}
\ee
where
\be
\begin{aligned}
&
B(\Omega)=X\cos\left(\frac{\pi \Omega}{2X}\right) -\frac{\pi (\Omega-1)}{2} \sin\left(\frac{\pi\Omega}{2X}\right),
\\
&D(\Omega)=\cos\left(\frac{\pi\Omega}{2X}\right) ,
\\
&C(\Omega)= 1-\Omega- \frac{X}{\pi} \sin\left[\frac{\pi\Omega}{X}\right].
\end{aligned}
\ee
One can check that for small $X$ and $\Omega \approx 1$  functions $B$ and $D$ are proportional to $\Sigma_{b,d}$:
$$B \propto \Sigma_{\rm b} (\Gamma=0), \quad D \propto \Sigma_{\rm d} (\Gamma=0).$$
 Exactly at frequencies, corresponding to zeroes of $B$ or $D,$
 the current is given by 
$1/(2 C).$ This allows one to find envelopes of bright and dark super-resonant peaks. For 
dark modes, condition $D(\Omega)=\cos(\pi \Omega/2X)=0,$ yields $\sin(\pi \Omega/X)=0,$ and consequently, $1/2C = 1/2(1-\Omega). $ For bright modes, condition $B(\Omega) =0 $ yields: $\cot(\pi \Omega/2X)=\pi(\Omega-1)/2 X. $ Expressing form this equation $\sin( \pi \Omega/X)$ and substituting to $1/2C,$ we finally arrive at the following formulas for envelopes
\be
\left( \frac{J_{\rm sup}^{n=1}}{J_0}\right)_{\rm env}\!\!\!\!\!\!=\left \{\begin{array}{l}
 \frac{1}{2(1-\Omega)} 
 \frac{4 X^2 + \pi^2 (1-\Omega)^2}{8 X^2 + \pi^2 (1-\Omega)^2}, \quad \rm {for~ bright~ modes}, 
 \\
 \\ \frac{1}{2(1-\Omega)},\quad \rm {for~ dark~ modes}. 
 \end{array}\right.
\label{eq-env-strong}
\ee
It is worth noting that difference between two envelopes becomes visible only for very small $X$ as illustrated in Fig.~\ref{fig_strong_super}c. 

 Equation~\eqref{eq-super-2} describes vicinity of the  fundamental resonance. Let us now assume that $X \lesssim \Omega \ll 1.$ To this end, we make several steps. First, we replace $\Omega \to t \Omega,~X \to t X, ~\Gamma \to \Gamma t$ and take limit $t \to 0.$ Thus obtained equation depends on parameter $z = \cot \left[{\pi \sqrt{\Omega } \sqrt{\Omega +i \Gamma }}/{(2 X)}\right]\approx \cot \left[{\pi \left(\Omega +{i \Gamma
 }/{2}\right)}/{(2 X)}\right].$ Next step is to make again rescaling $X \to t X, ~\Gamma \to \Gamma t$ and take limit $t \to 0$ keeping parameter $z$ untouched. Finally, one has to expand numerator and denominator of the obtained equation up to second order with respect to $\Gamma.$ Then, we get
\be
\frac{J_{\rm sup}}{J_0}\approx \frac{ \Gamma^2}{\Gamma^2+ \left({2 X}/{\pi}\right)^2 \sin^2(\pi \Omega/X)},\quad {\rm for}\quad X \lesssim \Omega \ll 1.
\label{eq-Gamma-to-zero}
\ee
Although we assume in this section, $\Gamma \ll X, $ we keep term $\Gamma^2$ in the denominator to regularize singularities at the frequencies corresponding to $\sin(\pi \Omega/X)=0.$ We also notice that for $\Gamma \gg X$ Eq.~\eqref{eq-Gamma-to-zero} tends to unity thus matching Eq.~\eqref{reson1} for $\Omega \gg \Gamma. $

To finish discussing the case $\Gamma \ll X,$ we consider the limit of very small frequency. Substituting in Eq.~\eqref{Eq_J} $ \Omega \to \Omega t$ and $\Gamma \to t \Gamma,$ and taking the limit $t \to 0,$ we get the Drude peak 
\be
\frac{J_{\rm sup}}{J_0}\approx \frac{ \Gamma^2}{\Gamma^2+ \Omega^2},\quad {\rm for}\quad X \lesssim \Omega \sim \Gamma.
\label{eq-Drude1}
\ee
in agreement with Eq.~\eqref{Jdrude} taken at $s_2 \ll s_1$ ($X\ll 1$).

\section{ Super-resonant regime with intermediate coupling}
\begin{figure}[h!]
\includegraphics[width=0.41
\textwidth]{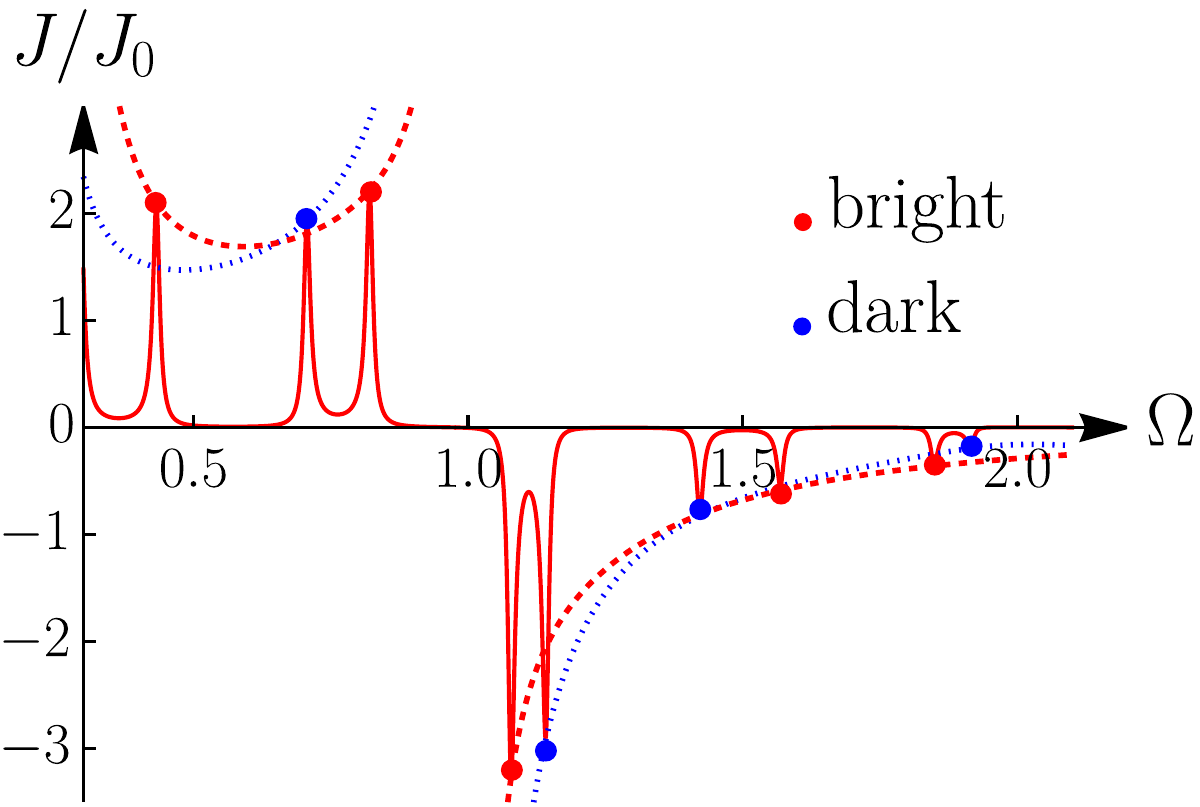}
\caption{Bright and dark peaks for intermediate coupling, $X=0.23$ and small relaxation rate, $\Gamma=0.015,$ corresponding to super-resonant regime. Thick line -- photo-response described by Eq.~\eqref{Eq_J}, red dashed line -- envelope of the bright modes, blue dashed line -- envelope of the dark modes. 
} \label{fig_envelopes}
\end{figure}

 In this section we briefly consider the case $\Gamma \ll 1$ and $X \sim 1$ corresponding to super-resonant regime with intermediate coupling strength. Typical response is shown in Fig.~\ref{fig_envelopes}. 
Analytical expression, Eq.~\eqref{Eq_J}, can not be essentially simplified in this case. Here, we discuss how to get analytical expressions for envelopes of bright and dark modes that generalize Eq.~\eqref{eq-env-strong} for the case of arbitrary $X$ and $\Omega$ and very small $\Gamma.$ We will use method previously used in Ref.~\cite{Gorbenko2024}.

As we mentioned in the beginning of the paper, peaks in the response corresponds to zeroes of $\Sigma_{\rm b}$ and $\Sigma_{\rm d} $
at $\Gamma=0.$
In the dimensionless form these equations yield 
\be
\begin{aligned}
&\Sigma_{\rm b}(\Gamma=0)=0 ~ \Leftrightarrow ~ 
X \cot \left(\frac{\pi \Omega }{2 X}\right)+\cot \left(\frac{\pi \Omega }{2}\right)=0,
\\
&\Sigma_{\rm d}(\Gamma=0)=0 ~ \Leftrightarrow ~\cot \left(\frac{\pi \Omega }{2 X}\right)+ X \cot \left(\frac{\pi \Omega }{2}\right)=0.
\end{aligned}
\ee
From these equations we find that $ \cot \left({\pi \Omega }/{2 X}\right)=-\cot \left({\pi \Omega }/{2}\right)/X$ at the positions of bright peaks and $ \cot \left({\pi \Omega }/{2 X}\right)=- X \cot \left({\pi \Omega }/{2}\right)$ at the positions of the dark peaks. 
Next, we use these formulas in  linear-in-$\Gamma$ expansion of  trigonometric functions entering exact formula \eqref{Eq_J} (see also formulas in Appendix~\ref{App-B-current}):
 \begin{align}
 &\cot \left(\frac{\pi \sqrt{\Omega (\Omega +i \Gamma )}}{2 X}\right)\approx \cot \left(\frac{\pi \Omega }{2 X}\right)-\frac{i \pi 
 \Gamma }{4 X \sin ^2\left(\frac{\pi \Omega }{2 X}\right)}
\\
 \nonumber
 &= \left\{ \begin{array}{l}
 -\frac{\cot \left(\frac{\pi \Omega }{2}\right)}{X}
-\frac{i \pi \Gamma }{4 X} \left[\frac{\cot ^2\left(\frac{\pi \Omega }{2}\right)}{X^2}+1\right],~\rm{for~bright~modes, } 
 \\
 \\
 \!\! -X \cot \!\left(\frac{\pi \Omega }{2}\right)-\frac{i \pi \Gamma \left[X^2 \cot ^2\left(\frac{\pi \Omega }{2}\right)\!+\!1\right]}{4 X}\!,~\rm{for~dark~modes. }
 \end{array} \right. 
\end{align}
Substituting these equations into exact equation \eqref{Eq_J} and taking the limit $\Gamma \to 0,$ we arrive at envelopes of both types of modes. Corresponding formulas are rather cumbersome and we present them in Appendix \ref{App-env} (see Eq.~\eqref{eq_envelopes_general}). Corresponding envelopes are shown in Fig.~
\ref{fig_envelopes} by dashed lines. For  $X \ll 1$ and $|\Omega - 1| \ll 1,$ Eqs.~\eqref{eq_envelopes_general} simplify to Eqs.~\eqref{eq-env-strong}. 

\section{Current control by gate voltages} \label{sec-control}
In the previous sections, we studied dependence of the photocurrent on the excitation frequency. In experiments, however, it is much simpler to fix the source frequency and vary the gate voltage. The goal of this section is to study the dependence on the parameter $X=s_2/s_1,$ which can be easily controlled by gates and discuss a general phase diagram in the $(X,\Omega)$ plane  focusing on gate-controlled transition from super-resonant regime to resonant regime with decreasing coupling strength. 

\begin{figure}[h!]
\includegraphics[width=0.5 \textwidth]{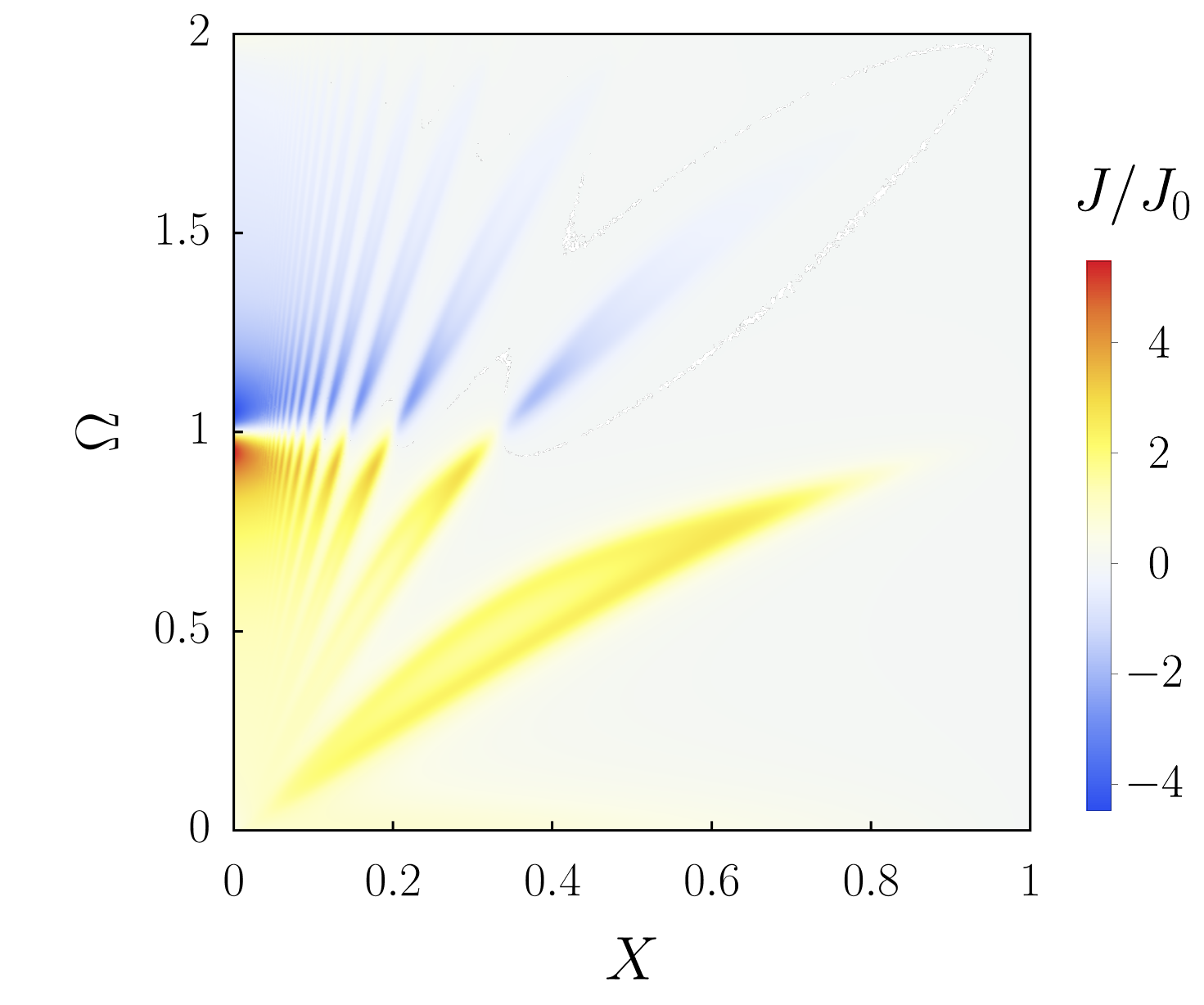}
\caption{ 
Heatmap of the current plotted with the use of    exact formula, Eq.~\eqref{Eq_J} as a function of the parameter $X=s_2/s_1,$ describing depth of the density modulation,  and dimensionless frequency $\Omega=\omega/\omega_0.$   for  $\Gamma=\gamma/\omega_0=0.1.$ 
}
\label{fig_transition}
\end{figure}

\begin{figure}[h!]
\includegraphics[width=0.5 \textwidth]{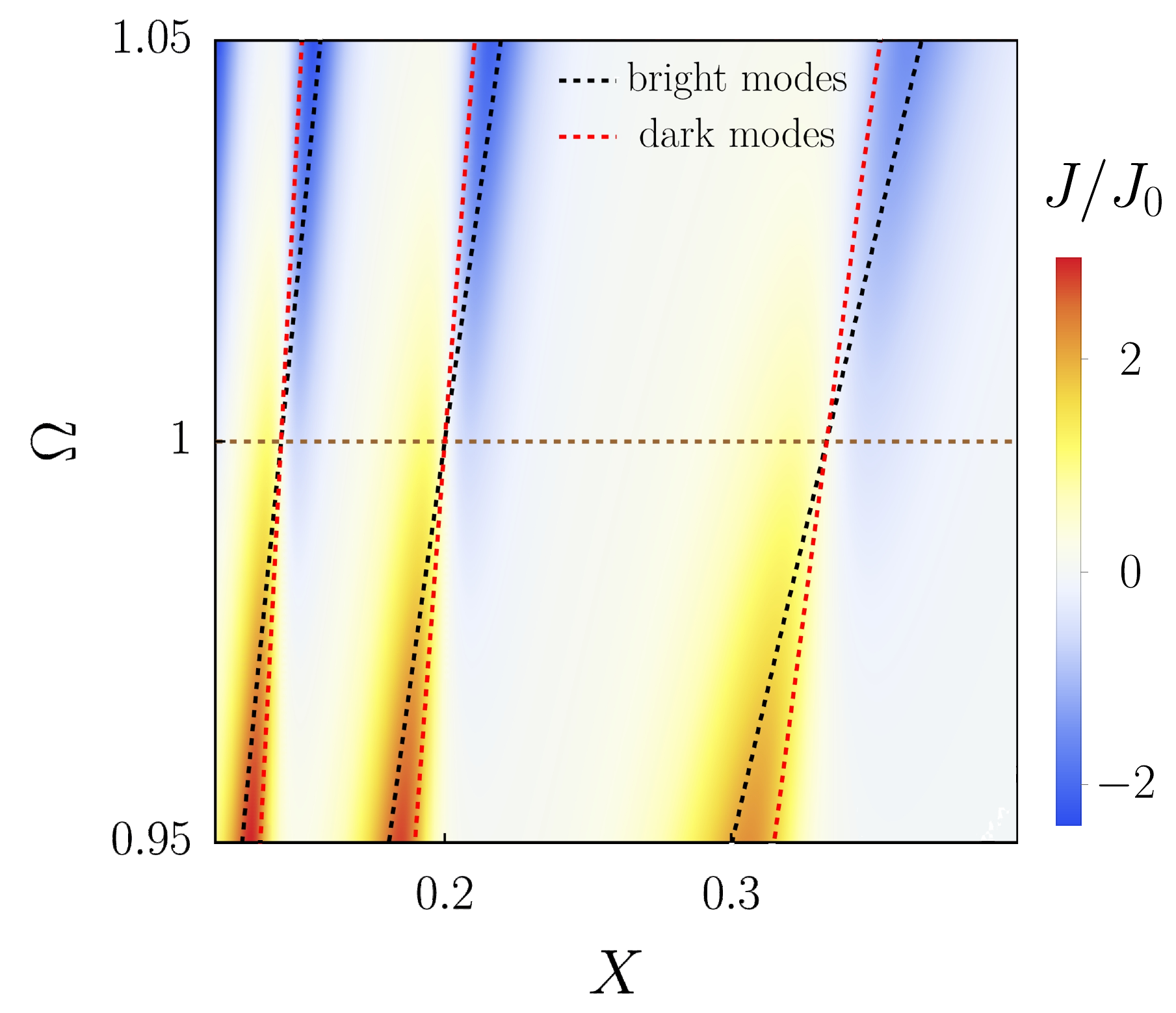}
\caption{ 
Heatmap of the exact expression, Eq.~\eqref{Eq_J} for current in the vicinity of $\Omega = 1$  for $\Gamma  = 0.1$. Dashed lines show position of the bright and dark modes.}
\label{fig_Omega_1}
\end{figure}

Figure~\ref{fig_transition} shows heat map of the photocurrent on the plane $(X,\Omega)$ plotted by using Eq.~\eqref{Eq_J}. The response is positive for small frequency, and negative for  higher frequencies, $1<\Omega<2$ in accordance with Figs.~\ref{fig_strong_resonant} and \ref{fig_strong_super}. Region  $\Omega \approx 1$ deserves a special attention. We plot this region in more detail in  Fig.~\ref{fig_Omega_1}. By dashed lines  we show positions of the  bright and dark modes, which are found by solution of Eq.~\eqref{sigma_bd_0} at $\Gamma=0.$  As seen, frequencies of  bright and dark modes  coincide at $\Omega=1$ (see a more detailed discussion of degeneracy conditions for  two types of modes in Ref.~\cite{Gorbenko2024}). Near the degeneracy point, effects related to finite  $\Gamma$ comes into play. As a result response can change sign for fixed $\Omega$ with increasing $X,$ in particular, along dashed horizontal line. 

\begin{figure}[h!]
\includegraphics[width=0.5 \textwidth]{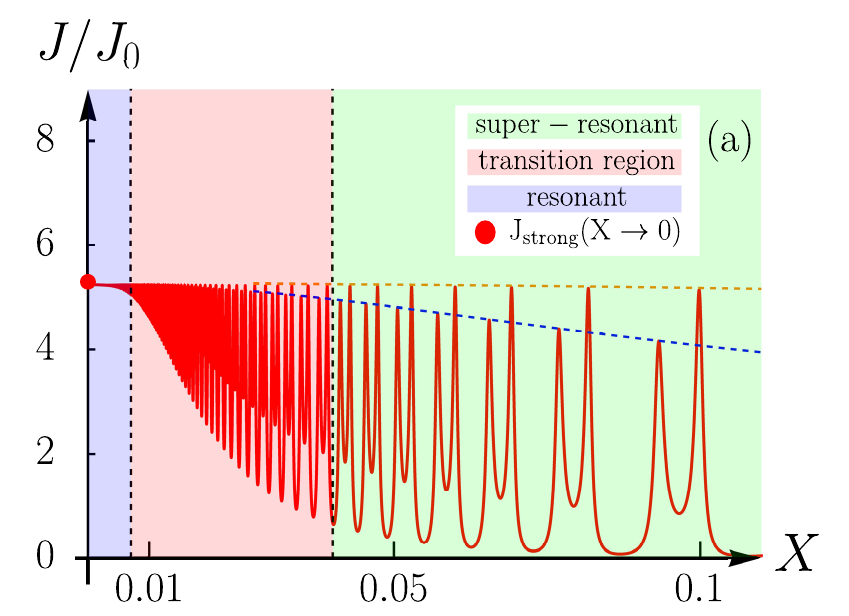}
\includegraphics[width=0.5 \textwidth]{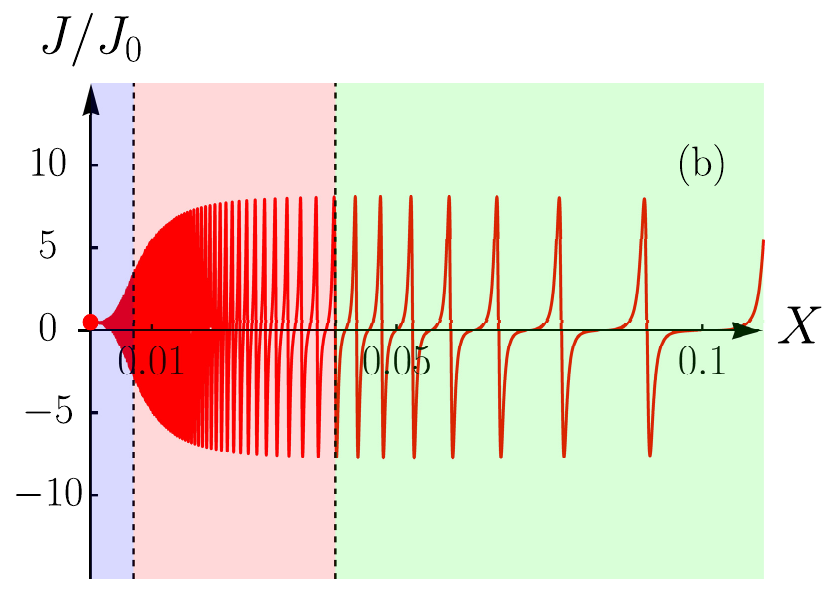}
\includegraphics[width=0.5 \textwidth]{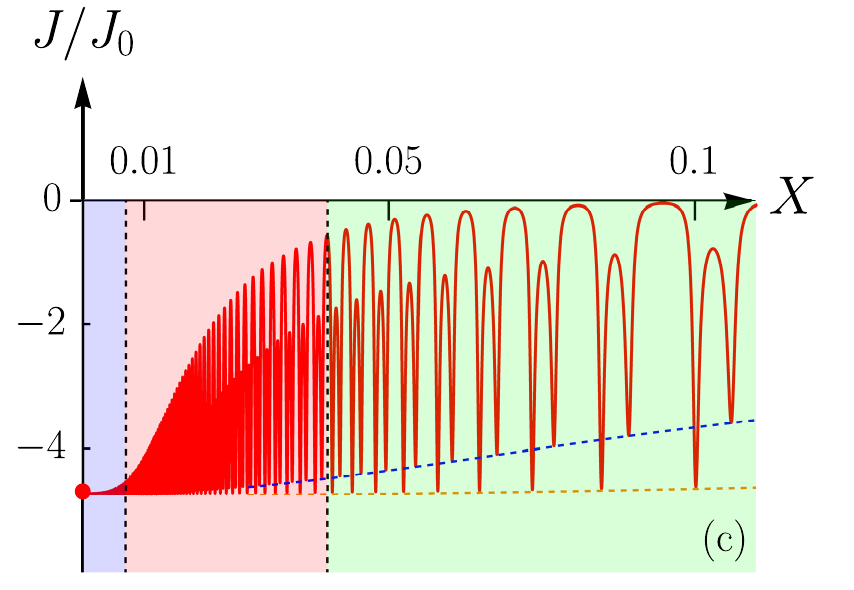}
\caption{ 
Dependence of current on $X$ for different values of $\Omega$ for $\Gamma=  0.013$: $\Omega=0.9$ (a), $\Omega=1$ (b),  $\Omega=1.1$ (c).  In all cases, system shows transitions to the resonant regime with decreasing $X.$ Thick red dots on the vertical axes corresponds to Eq.~\eqref{eq-res}. Dashed blue and  orange  lines in panel (a) and (c) show envelopes of bright and dark modes, respectively. Splitting of bright and dark modes is absent in  figure (b) due to degeneracy in the point $\Omega=1$ (see  Fig.~\ref{fig_Omega_1})    }
\label{fig_transition_BD}
\end{figure}

To further illustrate the possibilities of response control by  using gate voltages, we present  Fig.~\ref{fig_transition_BD}, which shows  $X-$dependence of the response at $\Omega<1,~\Omega=1$ and $\Omega>1$ (panels (a), (b), and (c), respectively).  Let us list most important features of this figure.  First, we see that the response oscillates but remains positive for $\Omega<1$ and negative for $\Omega>1 $     in accordance with Fig.~\ref{fig_transition}. At $\Omega \approx 1$ the response oscillates and change sign in agreement with  Fig.~\ref{fig_Omega_1}. Secondly, all three panels in  Fig.~\ref{fig_transition_BD} show transition from the resonant  regime to the  super-resonant regime with decreasing $X.$ Indeed, for large $X$ one can see well separated super-resonant peaks, which start to overlap with decreasing $X$ and finally, at $X \to 0$ tend to the $\Omega-$dependent value given by Eq.~\eqref{eq-res}. Finally, we notice that splitting into bright and dark modes is seen at $\Omega<1$ and $\Omega>1 $ but is  not seen  at  $\Omega=1$ because of mode degeneracy illustrated in Fig.~\ref{fig_Omega_1}.          

\section{Summary}
\label{sec-summary}

We have developed a non-perturbative theory of the plasmonic ratchet effect in a tunable lateral plasmonic crystal (LPC). Our approach fully accounts for arbitrarily strong modulation of the electron density by a dual grating gate, going beyond the conventional minimal model that relies on a perturbative treatment of both the radiation field and the periodic static potential. In the present theory, the static potential that creates the LPC -- via alternating regions with plasma wave velocities \( s_1 \) and \( s_2 \) -- is treated exactly, while the optical excitation is still considered within a perturbative framework.

The key results of our work are as follows:

\begin{itemize}
 \item 
 In asymmetric structures exhibiting the ratchet effect, the electromagnetic field excites two types of modes, the so-called ``dark'' and ``bright'' modes, with amplitudes of comparable magnitude. Unlike in the perturbative regime, where these modes are degenerate and indistinguishable, non-perturbative effects eliminate the degeneracy, allowing these two types of modes to interfere non-trivially.

 \item With increasing the coupling strength, controlled by depth of the density modulation, a giant enhancement of the photo-response at the fundamental plasmonic resonance at frequency $\omega_0$ occurs due to coherent interference between bright and dark modes. This results in a photocurrent that can exceed the low-frequency Drude peak by a factor scaling as $\sim \omega_0^2 \Delta^2 / \gamma^2$, where $\Delta = (s_1 - s_2)/s_1$ and $\gamma$ is the momentum relaxation rate. This factor can reach 10-100 in the high-quality structures.

 \item The theory describes the transition from the weak-coupling regime ($\Delta \ll 1$) to the strong-coupling regime ($X= 1-\Delta=s_2 / s_1 \ll 1$). In the strong-coupling limit, the system exhibits two distinct operational regimes controlled by the damping rate: a \emph{resonant} regime (for $\gamma \gg s_2/L$) characterized by a strongly asymmetric peak with the shape of the Fano type at fundamental frequency $\omega_0$
 as well as smaller peaks at multiples of $\omega_0$, and a \emph{super-resonant} regime (for $\gamma \ll s_2/L$) where a dense spectrum of fine peaks emerges due to the interplay of plasma waves in both the active (high-conductivity) and passive (low-conductivity) regions.

 \item The photocurrent can be efficiently controlled by the gate voltages, which tune the coupling parameter \(\Delta\). At a fixed frequency near resonance, the current oscillates in sign as a function of the ratio \(s_2/s_1\), providing a powerful electrical means for signal manipulation.
\end{itemize}

Our work establishes a comprehensive theoretical framework for the ratchet effect in gate-defined plasmonic crystals. It explains the giant resonant response, the role of mode interference, and the transition between resonant and super-resonant behaviors observed in recent experiments on transmission of THz signal through lateral plasmonic crystal \cite{Sai2023}. The theory provides a foundation for the design of tunable, high-efficiency THz detectors and frequency-selective current sources based on plasmonic ratchets.

\section*{Acknowledgments}
The study was supported by the 
Russian Science Foundation under grant 24-62-00010. The work of I.G. and S.P. was also partially supported by the Theoretical Physics and Mathematics Advancement Foundation ``BASIS''.

\appendix
\begin{widetext}

\section{Calculation details} \label{AppHD}
\subsection{Transfer matrix formalism}

For $\gamma \neq 0$ decay of the plasmonic oscillations due to momentum scattering can be compensated by the energy gain from the THz radiation. In this case, the only solution which is finite for $x \to \pm \infty$ does not depend on the cell number $M$. Skipping index $M$ and 
introducing vectors 
\be
\Psi_1 = 
\begin{pmatrix}
 A_1 \\ 
 B_1 
\end{pmatrix},
\Psi_2 = 
\begin{pmatrix}
 A_2 \\ 
 B_2 
\end{pmatrix}
\ee 
we arrive at a system of coupled equation determining the solution which is finite for $|x| \to \infty :$ 
\be
\begin{aligned}
&\Psi_1 = (1-\hat{T}_{21} \hat{T}_{12})^{-1} (\hat{T}_{21} f_2 + f_1), \\
&\Psi_2 = (1-\hat{T}_{12} \hat{T}_{21})^{-1} (\hat{T}_{12} f_1 + f_2).
\end{aligned}
\label{Eq-T-f}
\ee
Here, transfer matrices $\hat{T}_{12},~ \hat{T}_{21}$ and vectors $f_1,~ f_2$ are found using boundary conditions Eqs.~\eqref{Eq-BC}:
\be
\hat{T}_{12} = \frac{s_1^2}{2 q_1 s_2^2}
\begin{pmatrix}
 e^{i (q_1-q_2) L_1}(q_1 + q_2) & e^{-i (q_1+q_2) L_1}(q_1 - q_2) \\
 e^{i (q_1+q_2) L_1}(q_1 - q_2) & e^{-i (q_1-q_2) L_1}(q_1 + q_2)
\end{pmatrix}
,
\ee
\be
\hat{T}_{21} = \frac{s_2^2}{2 q_2 s_1^2}
\begin{pmatrix}
 e^{i q_2 (L_1+L_2)}(q_1 + q_2) & e^{-i q_2 (L_1+L_2)}(q_1 - q_2) \\
 e^{i q_2 (L_1+L_2)}(q_1 - q_2) & e^{-i q_2 (L_1+L_2)}(q_1 + q_2)
\end{pmatrix}
,
\ee
\be
f_1 = \frac{i F_0 (q_1^2 - q_2^2)}{4 m q_1 q_2^2 s_1^2} \left[\begin{pmatrix}
 -1 \\
 1
\end{pmatrix} 
+ \frac{h q_1^2 q_2^2}{(q_1^2-k^2)(q_2^2-k^2)} \left[
\cos{\phi} \begin{pmatrix}
 -1 \\
 1
\end{pmatrix}
-\frac{i k \sin{\phi}}{q_1} \begin{pmatrix}
 1 \\
 1
\end{pmatrix} 
\right]
\right] ,
\ee
\be
f_2 = \frac{i F_0 (q_1^2 - q_2^2)}{4 m q_1^2 q_2 s_2} \left[\begin{pmatrix}
 -e^{-i q_2 L_1} \\
 e^{i q_2 L_1}
\end{pmatrix} 
+ \frac{h q_1^2 q_2^2}{(q_1^2-k^2)(q_2^2-k^2)} \left[
\cos{(k L_1 + \phi)} \begin{pmatrix}
 -e^{-i q_2 L_1} \\
 e^{i q_2 L_1}
\end{pmatrix}
-\frac{i k \sin{(k L_1 + \phi)}}{q_2} \begin{pmatrix}
 e^{-i q_2 L_1} \\
 e^{i q_2 L_1}
\end{pmatrix} 
\right]
\right] .
\ee

\subsection{Amplitudes $A_{1,2}$,$B_{1,2}$}\label{App-AB12bd}

In the general case of arbitrary $L_1$ and $L_2,$ analytical expressions for $A_{1,2}$ and $B_{1,2}$ are quite cumbersome. Here we present them assuming $L_1 = L_2 = L/2.$ For this case, solution of Eq.~\eqref{Eq-T-f} can be simplified and presented in a form of Eq.\eqref{Eq_AB_bd} with the coefficients $A_{1,2}^{\rm b,d}$ given by:
\begin{align}
&A^{\rm b}_1 = e^{-i \frac{q_1 L}{4}} \frac{i F_0 (q_1^2-q_2^2) \csc{\frac{q_1 L}{4}}}{4 m q_1 q_2^2 s_1} \left(1-\frac{h k q_1^2 q_2 \cot{\frac{q_2 L}{4}} \sin{\phi}}{(q_1^2-k^2)(q_2^2-k^2)} \right),
\\
&A^{\rm d}_1 = - e^{-i \frac{q_1 L}{4}} \frac{F_0 h q_1 (q_1^2-q_2^2) \csc{\frac{q_1 L}{4}} \cot{\frac{q_2 L}{4}} \cos{\phi} }{4 m s_1 (q_1^2-k^2)(q_2^2-k^2)},
\\
&B^{\rm b}_1 = - e^{i \frac{q_1 L}{2}} A^{\rm b}_1, \, \, B^{\rm d}_1 = e^{i \frac{q_1 L}{2}} A^{\rm d}_1
\label{Eq_AB_bd_1}
\\
&A^{\rm b}_2 = -e^{-i \frac{3 q_2 L}{4}} \frac{i F_0 (q_1^2-q_2^2) \csc{\frac{q_2 L}{4}}}{4 m q_1^2 q_2 s_2} \left(1+\frac{h k q_1 q_2^2 \cot{\frac{q_1 L}{4}} \sin{\phi}}{(q_1^2-k^2)(q_2^2-k^2)} \right),
\\
&
A^{\rm d}_2 = -e^{-i \frac{3 q_2 L}{4}} \frac{F_0 h q_2 (q_1^2-q_2^2) \csc{\frac{q_2 L}{4}} \cot{\frac{q_1 L}{4}} \cos{\phi} }{4 m s_2 (q_1^2-k^2)(q_2^2-k^2)},
\\
&
B^{\rm b}_2 = - e^{i \frac{3 q_2 L}{2}} A^{\rm b}_2, \, \, B^{\rm d}_2 = e^{i \frac{3 q_2 L}{2}} A^{\rm d}_2.
\label{Eq_AB_bd_2}
\end{align}

\section{Classification of relevant contributions to the dc response}\label{App-relevant}
In this Appendix, we analyze different contributions to the ratchet current and explain how to chose four relevant terms entering Eq.~\eqref{Eq_J}.

First, we rewrite oscillating concentration  $n(x) = n_\alpha(x)$ and oscillating velocity  $v(x) = v_\alpha(x)$ 
in a schematic way, separating terms of different physical sense (we skip index $\alpha$):   
\be
\begin{aligned}
&n(x) = n_0^{\rm b}+n^{\rm b}+ n^{\rm d}+ n^{\rm ext},
\\
&
v(x)= v_0^{\rm b}+v^{\rm b}+v^{\rm d}+v_{0}^{\rm ext}+ v^{\rm ext}.
\label{Eq-contrib}
\end{aligned}
\ee
Here index ``0'' stands for ac responses with respect to homogeneous component of the external electric field.    
% Here,  we introduced inhomogeneous ``dark''  solutions $n^{\rm d}, v^{\rm d}$; ``bright'' solutions: homogeneous $n_0^{\rm b}, v_0^{\rm b}$ and inhomogeneous $n^{\rm b}, v^{\rm b}$; external homogeneous solution for velocity $v_0^{\rm ext}$ and inhomogeneous external solutions $n^{\rm ext}, v^{\rm ext}$. Then concentration and velocity can be rewritten as 
Substituting Eq.~\eqref{Eq-contrib} into 
Eq.\eqref{Eq-Jdc}, we get $4\times 5=20$ contributions. 

Let us classify these contributions.
Nine of these terms do not contain index ``0'' neither in $v$ not in $n$: $n^{\rm b} v^{\rm d}$, $n^{\rm d} v^{\rm b}$, $n^{\rm b} v^{\rm ext}$, $n^{\rm ext} v^{\rm b}$, $n^{\rm d} v^{\rm ext}$, $n^{\rm ext} v^{\rm d}$,  $n^{\rm ext} v^{\rm ext}$, $ n^{\rm b} v^{\rm b}  , \,   n^{\rm d} v^{\rm d}  .$ Therefore, they are proportional to $h^2$ and can be neglected.
Five terms  $n_0^{\rm b} v_0^{\rm b}, \, n_0^{\rm b} v^{\rm b}, \, n^{\rm b} v_0^{\rm b},  \,  (n_0^{\rm b}$, $n^{\rm b}) v^{\rm ext}_0$  are not small but given by  products of symmetrical and asymmetrical functions (with respect to $x=L_1/2=L/4$) and  vanished after integration over crystal cell. 
The rest six terms can be combined as follows: $j_0 \propto \left<n^{\rm ext} v^{\rm ext}_0\right>$, $j_{\rm b} \propto \left<n_0^{\rm b} v^{\rm ext}+n^{\rm ext} v_0^{\rm b}\right>$, $j_{\rm d} \propto \left<n^{\rm d} v^{\rm ext}_0\right>$ and $j_{\rm bd} \propto \left<n_0^{\rm b} v^{\rm d}+n^{\rm d} v_0^{\rm b}\right>$ (here $\langle \cdots\rangle$ stands for averaging over PC cell).  The latter terms are non-zero and proportional to $h.$ They  give four contributions enetering Eq.~\eqref{Eq_J}.

\section{ Analytical expression for the direct current } \label{App-B-current}

For $L_1=L_2= L/2,$ analytical expressions for four terms entering Eq.~\eqref{Eq_J} are given by

\begin{align}
 & \frac{j_{\rm bd}}{J_0}= \frac{\left( s_1^2-s_2^2 \right)^2 k^2 q_1^2 q_2^2}{4 (q_1^2-k^2) (q_2^2-k^2) (s_1^2+s_2^2) \Sigma_b^* \Sigma_d} \times\\
 \nonumber
 &\left( \cot{\frac{q_1 L}{4}} \left[\frac{s_1^2 }{\omega^2+\gamma^2} \cot{\frac{q_2^* L}{4}} - \frac{\left(s_1^2-s_2^2\right) }{\omega \sqrt{\omega^2 + \gamma^2}} \cot{\frac{q_2 L}{4}} \right] -\frac{s_2^2 }{\omega^2+\gamma^2} \cot{\frac{q_1^* L}{4}}\cot{\frac{q_2 L}{4}} \right),
\\
 &\frac{j_{\rm b}}{J_0}=-\frac{i \gamma k^2 (q_1^2-q_2^2) s_1^2 s_2^2 \omega}{4 q_1^2 q_2^2 (s_1^2+s_2^2) \Sigma_{\rm b}} \left(\frac{(k^2+q_1^2) \cot{\frac{q_1 L}{4}}}{|q_1^2-k^2|^2 s_1^3}+\frac{(k^2+q_2^2) \cot{\frac{q_2 L}{4}}}{|q_2^2-k^2|^2 s_2^3}\right),
\\
%\end{align}
%\begin{align}
 &\frac{j_{\rm d}}{J_0}= -\frac{i \gamma k^2 (q_1^2-q_2^2) s_1 s_2 \Sigma_{\rm b}}{4 (q_1^2-k^2) (q_2^2-k^2) (s_1^2+s_2^2) (\omega-i \gamma) \Sigma_{\rm d}},
%\end{align}
%\begin{align}
\\
 &\frac{j_{0}}{J_0}= -\frac{i \gamma k^2 s_1^2 s_2^2}{4 (s_1^2+s_2^2) (\omega - i \gamma)} \left[\frac{1}{s_1^2 (q_1^2- k^2)}-\frac{1}{s_2^2(q_2^2-k^2)}\right ].
\end{align}

\section{Envelopes of the bright and dark modes}\label{App-env}
Here, we present analytical equations for the envelopes of the bright and dark modes 
in the super-resonant regime valid for any frequency $\Omega$ and arbitrary coupling strength, $0<X<1$: 
\be
\begin{aligned}
&\frac{J_{\rm env}^{\rm bright}}{J_0}=\frac{\left(1-X^2\right) \Omega ^2 \left(2 X^4-\left(X^2-1\right) \cos (\pi \Omega )+X^2+1\right)}{\left(X^2+1\right) \left(\Omega
 ^2-1\right) \left(X^2-\Omega ^2\right) \left(\left(1-X^2\right) \cos (\pi \Omega )+3 X^2+1\right)}
 \\
 &+\frac{2 X^2
 \left(1-X^2\right)^2 \left(X^4-2 \left(X^2+1\right) \Omega ^4+X^2+\left(X^4-X^2+1\right) \Omega ^2+\Omega ^6\right) \sin (\pi 
 \Omega )}{\pi \left(X^2+1\right) \Omega \left(\Omega ^2-1\right)^2 \left(X^2-\Omega ^2\right)^2 \left(\left(1-X^2\right) \cos
 (\pi \Omega )+3 X^2+1\right)},
 \\
 &\frac{J_{\rm env}^{\rm dark}}{J_0}=\frac{\left(1-X^2\right) \Omega \left(-2 X^2 \sin (\pi \Omega )+\pi \left(2 X^4+X^2+1\right) \Omega +\pi \Omega \cos (\pi \Omega
 )\right)}{\pi \left(X^2+1\right) \left(\Omega ^2-1\right) \left(X^2-\Omega ^2\right) \left(\left(X^2-1\right) \cos (\pi \Omega
 )+3 X^2+1\right)}.
 \end{aligned}
 \label{eq_envelopes_general}
 \ee
These equations do not take into account some additional peaks that appears for special values of $X,$ corresponding to degeneracy of the bright and dark frequencies. One can check that in the limit of strong coupling, $X \ll 1$, Eq.~\eqref{eq_envelopes_general} simplifies to Eq.~\eqref{eq-env-strong} 

\end{widetext}
\bibliography{main}

\end{document}